\documentclass[twocolumn]{aastex631}

\usepackage{savesym}
\usepackage{tipa} 
\savesymbol{tablenum}
\usepackage{siunitx}
\restoresymbol{SIX}{tablenum}
\DeclareSIUnit{\jansky}{Jy}
\DeclareSIUnit{\MSPS}{MSPS}
\DeclareSIUnit{\byte}{B}
\DeclareSIUnit{\tecu}{TECu}
\DeclareSIUnit{\bit}{bit}
\DeclareSIUnit{\sample}{S}
\DeclareSIUnit{\dmunit}{pc~cm^{-3}}
\DeclareSIUnit{\millisec}{ms}
\newcommand{\kkoname}{k'ni\textipa{P}atn k'l$\left._\mathrm{\smile}\right.$stk'masqt}
\newcommand{\rmtec}{\mathrm{TEC}}

\usepackage{cancel}
\usepackage{aas_macros}

\usepackage{amsmath, amssymb,commath}

\newcommand{\corrname}{\texttt{PyFX}}

\newcommand{\sfxc}{\texttt{SFXC}}
\newcommand{\difx}{\texttt{DiFX}}

\newcommand{\nhat}{\widehat{\textbf{n}}}
\newcommand{\hdf}{\texttt{hdf5}}
\newcommand{\vdif}{\texttt{VDIF}}

\newcommand{\coda}{\texttt{coda}}
\newcommand{\kdm}{K_{\textrm{DM}}}
\DeclareSIUnit{\parsec}{pc}
\newcommand{\chimedatb}{\SI{19.3996}{\millisec}} 

\received{\today}
\revised{\today}
\submitjournal{AJ}
\shorttitle{Python-based VLBI Correlator}
\shortauthors{Leung et al.}

\begin{document}
\title{A VLBI Software Correlator for Fast Radio Transients}
\author[0000-0002-4209-7408]{Calvin Leung}
  \affiliation{Department of Astronomy, University of California Berkeley, Berkeley, CA 94720, USA}
  \affiliation{MIT Kavli Institute for Astrophysics and Space Research, Massachusetts Institute of Technology, 77 Massachusetts Ave, Cambridge, MA 02139, USA}
  \affiliation{Department of Physics, Massachusetts Institute of Technology, 77 Massachusetts Ave, Cambridge, MA 02139, USA}
  \affiliation{NASA Hubble Fellowship Program~(NHFP) Einstein Fellow}
\author[0000-0002-3980-815X]{Shion Andrew}
  \affiliation{MIT Kavli Institute for Astrophysics and Space Research, Massachusetts Institute of Technology, 77 Massachusetts Ave, Cambridge, MA 02139, USA}
\author[0000-0002-4279-6946]{Kiyoshi W.~Masui}
  \affiliation{MIT Kavli Institute for Astrophysics and Space Research, Massachusetts Institute of Technology, 77 Massachusetts Ave, Cambridge, MA 02139, USA}
  \affiliation{Department of Physics, Massachusetts Institute of Technology, 77 Massachusetts Ave, Cambridge, MA 02139, USA}
\author[0000-0002-1800-8233]{Charanjot Brar}
  \affiliation{Department of Physics, McGill University, 3600 rue University, Montr\'eal, QC H3A 2T8, Canada}
  \affiliation{Trottier Space Institute, McGill University, 3550 rue University, Montr\'eal, QC H3A 2A7, Canada}
\author[0000-0003-2047-5276]{Tomas Cassanelli}
  \affiliation{Department of Electrical Engineering, Universidad de Chile, Av. Tupper 2007, Santiago 8370451, Chile}
\author[0000-0002-2878-1502]{Shami Chatterjee}
  \affiliation{Cornell Center for Astrophysics and Planetary Science, Cornell University, Ithaca, NY 14853, USA}
\author[0000-0001-9345-0307]{Victoria Kaspi}
  \affiliation{Department of Physics, McGill University, 3600 rue University, Montr\'eal, QC H3A 2T8, Canada}
  \affiliation{Trottier Space Institute, McGill University, 3550 rue University, Montr\'eal, QC H3A 2A7, Canada}
\author[0009-0005-7115-3447]{Kholoud Khairy}
  \affiliation{Lane Department of Computer Science and Electrical Engineering, 1220 Evansdale Drive, PO Box 6109, Morgantown, WV 26506, USA}
  \affiliation{Center for Gravitational Waves and Cosmology, West Virginia University, Chestnut Ridge Research Building, Morgantown, WV 26505, USA}
\author[0000-0003-2116-3573]{Adam E.~Lanman}
  \affiliation{MIT Kavli Institute for Astrophysics and Space Research, Massachusetts Institute of Technology, 77 Massachusetts Ave, Cambridge, MA 02139, USA}
  \affiliation{Department of Physics, Massachusetts Institute of Technology, 77 Massachusetts Ave, Cambridge, MA 02139, USA}
  \affiliation{Department of Physics, McGill University, 3600 rue University, Montr\'eal, QC H3A 2T8, Canada}
  \affiliation{Trottier Space Institute, McGill University, 3550 rue University, Montr\'eal, QC H3A 2A7, Canada}
\author[0000-0002-5857-4264]{Mattias Lazda}
  \affiliation{Dunlap Institute for Astronomy \& Astrophysics, University of Toronto, 50 St.~George Street, Toronto, ON M5S 3H4, Canada}
  \affiliation{David A.~Dunlap Department of Astronomy \& Astrophysics, University of Toronto, 50 St.~George Street, Toronto, ON M5S 3H4, Canada}
\author[0000-0002-0772-9326]{Juan Mena-Parra}
  \affiliation{Dunlap Institute for Astronomy \& Astrophysics, University of Toronto, 50 St.~George Street, Toronto, ON M5S 3H4, Canada}
  \affiliation{David A.~Dunlap Department of Astronomy \& Astrophysics, University of Toronto, 50 St.~George Street, Toronto, ON M5S 3H4, Canada}
\author[0000-0002-5254-243X]{Gavin Noble}
  \affiliation{Dunlap Institute for Astronomy \& Astrophysics, University of Toronto, 50 St.~George Street, Toronto, ON M5S 3H4, Canada}
  \affiliation{David A.~Dunlap Department of Astronomy \& Astrophysics, University of Toronto, 50 St.~George Street, Toronto, ON M5S 3H4, Canada}
\author[0000-0002-8912-0732]{Aaron B.~Pearlman}
  \affiliation{Department of Physics, McGill University, 3600 rue University, Montr\'eal, QC H3A 2T8, Canada}
  \affiliation{Trottier Space Institute, McGill University, 3550 rue University, Montr\'eal, QC H3A 2A7, Canada}
  \affiliation{Banting Fellow}
  \affiliation{McGill Space Institute Fellow}
  \affiliation{FRQNT Postdoctoral Fellow}
\author[0000-0003-1842-6096]{Mubdi Rahman}
  \affiliation{Sidrat Research, 124 Merton Street, Suite 507, Toronto, ON M4S 2Z2, Canada}
\author[0000-0001-5504-229X]{Pranav Sanghavi}
  \affiliation{Department of Physics, Yale University, New Haven, CT 06520, USA}
\author[0000-0002-4823-1946]{Vishwangi Shah}
  \affiliation{Department of Physics, McGill University, 3600 rue University, Montr\'eal, QC H3A 2T8, Canada}
  \affiliation{Trottier Space Institute, McGill University, 3550 rue University, Montr\'eal, QC H3A 2A7, Canada}
\newcommand{\allacks}{
C. L. is supported by NASA through the NASA Hubble Fellowship grant HST-HF2-51536.001-A awarded by the Space Telescope Science Institute, which is operated by the Association of Universities for Research in Astronomy, Inc., under NASA contract NAS5-26555.
K.W.M. holds the Adam J. Burgasser Chair in Astrophysics and is supported by an NSF Grant (2008031).
V.\,M.\,K. holds the Lorne Trottier Chair in Astrophysics \& Cosmology, a Distinguished James McGill Professorship, and receives support from an NSERC Discovery grant (RGPIN 228738-13), from an R. Howard Webster Foundation Fellowship from CIFAR, and from the FRQNT CRAQ.
A.B.P. is a Banting Fellow, a McGill Space Institute~(MSI) Fellow, and a Fonds de Recherche du Quebec -- Nature et Technologies~(FRQNT) postdoctoral fellow.
}

\collaboration{99}{(CHIME/FRB Collaboration)}


\begin{abstract}
    One major goal in fast radio burst science is to detect fast radio bursts (FRBs) over a wide field of view without sacrificing the angular resolution required to pinpoint them to their host galaxies. Widefield detection and localization capabilities have already been demonstrated using connected-element interferometry; the CHIME/FRB Outriggers project will push this further using widefield cylindrical telescopes as widefield outriggers for very long baseline interferometry (VLBI). This paper
    describes an offline VLBI software correlator written in Python for the CHIME/FRB Outriggers project. It includes features well-suited to modern widefield instruments like multibeaming/multiple phase center correlation, pulse gating including coherent dedispersion, and a novel correlation algorithm based on the quadratic estimator formalism. This algorithm mitigates sensitivity loss which arises in instruments where the windowing and channelization is done outside the VLBI correlator at each
    station, which accounts for a 30 percent sensitivity drop away from the phase center. Our correlation algorithm recovers this sensitivity on both simulated and real data. As an end to end check of our software, we have written a preliminary pipeline for VLBI calibration and single-pulse localization, which we use in Lanman et al. (2024) to verify the astrometric accuracy of the CHIME/FRB Outriggers array.
\end{abstract}

\section{Introduction - VLBI and Fast Radio Bursts}

Very long baseline interferometry (VLBI) is a technique used to resolve spatial structures with the highest angular resolutions possible in astronomy. It relies on the phase-coherent recording of incident electric fields at widely-separated telescope stations, as well as stable timing precision at each telescope site. Since the 1960s, VLBI at ever-higher observing frequencies has pushed the angular resolution frontier to the microarcsecond level\citep{kellerman1988origin}. This has recently
delivering the world's first direct images of the environment around the supermassive black hole M87, enabling unique tests of general relativity and characterization of active galactic nuclei~\citep{collaboration2019first1}. 

A fundamental component in VLBI data analysis is the VLBI correlator, which reduces the raw voltage data into visibilities, from which spatial information can be gleaned. Correlator implementations have changed dramatically over the years since the early days of correlators running on custom-built computing hardware; modern FX correlators such as~\difx~run as parallelizable, modular software on generic computer clusters and place a strong
emphasis on flexibility. This enabled additional observational capabilities, including VLBI on transients, to be added flexibly to respond to observational demands. For example, \difx~supports VLBI on fast transient sources via features such as pulsar gating and dedispersion~\citep{deller2007difx}. These complex capabilities would have been difficult to add into custom-built hardware correlators after their initial design and construction.

An exciting frontier in radio interferometry is using VLBI to pinpoint, or localize, fast radio bursts (FRBs). FRBs are millisecond-duration, highly-dispersed radio transients now known to lie at cosmological distances~\citep{petroff2019fast}. They are of interest to the high-energy astrophysics community due to their extreme luminosities and diverse timescales (see, e.g.~\citep{nimmo2021burst,ryder2022probing}), and their strong connection to magnetars~\citep{collaboration2020bright,bochenek2020fast}. In addition, they are of growing interest to the cosmology community due to their
abundance~\citep{collaboration2021first2} and potential as cosmological tools through precise timing ~\citep{oguri2019strong,li2018strongly,leung2023wave} and extragalactic dispersion/Faraday rotation/scattering, which can probe the baryonic contents of our universe 
\citep{macquart2020census,mcquinn2014locating,prochaska2019probing,masui2015dispersion,ravi2019fast2,connor2022observed,madhavacheril2019cosmology}.

The one-off nature of FRBs means that localizing one-off FRBs requires widefield interferometric capabilities. In contrast, VLBI has traditionally been narrow-field science: observations are typically scheduled in advance and conducted on noteworthy targets using single-dish telescopes which must be 
coordinated and simultaneously pointed towards the target. Combining FRB
detections with astrometric VLBI capabilities in one instrument has therefore been a major challenge for the field. From the discovery of the first FRB~\citep{lorimer2007bright}, it took a decade before the first interferometric localization~\citep{chatterjee2017direct} pinpointed the FRB to a host galaxy, whose spectroscopic redshift~\citep{tendulkar2017host} conclusively established the extragalactic nature of the phenomenon. 

Since then, a select handful of bursts have been localized with VLBI, but almost every VLBI
localization has helped to uncover the origins of FRBs. These include the discovery of a persistent radio source in the local environment following the VLBI localization of FRB 20121102~\citep{marcote2017repeating}, the VLBI localization of FRB 20180916 to a progenitor slightly offset from a knot of star formation within its host galaxy~\citep{marcote2020repeating}, the localization of FRB 20190520 to a globular cluster in the nearby galaxy M81~\citep{kirsten2022repeating}, the localization of FRB
20201124A to a complex star-formating site in its host~\citep{nimmo2022milliarcsecond,xu2021fast}, and the surprising non-detection of persistent radio emission surrounding the highly-active repeating FRB 20220912A~\citep{hewitt2023milliarcsecond,feng2023extreme}. In the meantime, thousands of un-localized bursts have been detected, owing to the large collecting area, unique widefield reflectors, and fully-coherent beamforming capabilities of the FRB backend of the Canadian Hydrogen Intensity
Mapping Experiment (CHIME/FRB)~\citep{amiri2018chime}. 

CHIME/FRB Outriggers is a set of three telescopes which will improve on CHIME by combining its wide-field burst-finding capabilities with high-resolution astrometric VLBI techniques in a single VLBI array. The three outrigger telescopes will observe in tandem with CHIME, matching the field of view of CHIME. Each outrigger telescope has a cylindrical reflector with close-packed feeds. The first, the~\kkoname~Outrigger (hereafter, KKO) has 1/16 of the collecting area and number of feeds of CHIME (64 dual-polarization feeds; \SI{500}{\meter^2}), and is located $\SI{65}{\kilo\meter}$ from CHIME
in a green field, while the other two cylinders will have 1/8 the collecting area
and number of feeds (128 dual-polarization feeds; \SI{1000}{\meter^2}) of CHIME. The latter two stations will be located on existing observatory sites at Green Bank Observatory (GBO) and Hat Creek Radio Observatory (HCRO) respectively. The outriggers
will collectively be used as widefield VLBI stations to localize single pulses to 50 milliarcsecond precision. A preliminary but end-to-end demonstration of this capability using CHIME/FRB and a small testbed array of narrow-field VLBI outriggers was presented in~\citet{cassanelli2023fast}. Next generation FRB survey instruments like CHIME's sucessor CHORD~\citep{vanderlinde2019canadian} will rely on VLBI stations for burst localization. CHORD is not alone in this respect: FRB localizations
using VLBI are also a key goal of HIRAX and BURSTT, to name a few~\citep{crichton2022hydrogen,lin2022burstt}. 

In anticipation of new instruments coming online which will rely on VLBI to deliver FRB localizations, we present~\corrname~\footnote{\texttt{https://github.com/leungcalvin/pyfx-public}}: a dedicated VLBI software correlator which furthers the trend of more flexible, agile, and transparent correlation software. We hope that the transparency of~\corrname~can allow it to serve as a starting point for future FRB survey instruments counting on VLBI localizations to study FRBs at high angular resolution. We have written in Python given our anticipated computational
load, which will be modest compared to typical VLBI imaging observations with hundreds of baselines and many hours of integration time. Even so,~\corrname~supports modern features like multiple phase centers~\citep{morgan2011vlbi} and coherent dedispersion.
In the remainder of this paper we document the various parts of our VLBI correlator, which is currently being used for observing compact calibrators, pulsars, and FRBs. We first discuss the software frameworks used in our correlator (Section~\ref{sec:software}). Next, we review the method of compensating for delays in our data and dividing the overall correlation job into parallelizable subproblems as a function of frequency,
pointing, and time (Section~\ref{sec:chunking}). This takes into account coherent dedispersion in VLBI: a necessary feature for low-frequency observations of fast transients like pulsars and FRBs (Section~\ref{sec:gating}). Next, we describe how the post-channelized data are correlated in Section~\ref{sec:corr_algorithms}. We characterize the sensitivity loss resulting from the fact that our data are windowed by a polyphase filter bank (PFB) and channelized independently of our VLBI correlation
analysis, and mitigate this using a correlation algorithm based on the quadratic
estimator technique~\citep{tegmark1997how} which we have designed to mitigate PFB-related losses. We demonstrate this on both real and simulated data. We conclude by describing the current status of validating the correlator software (Section~\ref{sec:validation}) and introducing a candidate VLBI localization pipeline which we have applied to pulsars and FRBs localized to arcsecond scales on the CHIME-KKO baseline, where the in-beam calibration technique in~\citet{leung2021synoptic} has proven successful.

\section{Software Design and Frameworks}\label{sec:software}
~\corrname~is implemented almost entirely in Python 3. Nearly all computations are performed using commonly-used signal processing tools provided by~\texttt{numpy} and~\texttt{scipy}, and~\texttt{astropy}, except for the standalone delay model~\texttt{calc}/\texttt{difxcalc}~\citep{eubanks1991consensus,gordon2016difxcalc}. While performance is a secondary goal, certain components of~\corrname are optimized for speed. For instance, if GPUs are available on the computing environment, efficient
Pytorch FFTs can seamlessly replace the~\texttt{scipy} default.  

A major motivating factor for developing a new correlator instead of using existing software correlators, e.g. \difx~\citep{deller2007difx} and \sfxc~\citep{keimpema2015sfxc}, and \texttt{lcorr}~\citep{smits2017beamformer} arises from native support for the data format used by CHIME to store raw data. We use the term ``voltage'' data to refer to the raw electric field received at each antenna, and the term ``baseband'' data to refer to the
voltages after channelization.
CHIME data are channelized immediately after digitization, making the voltage data unavailable, but the CHIME correlator, called~\texttt{kotekan}~\citep{recnik2015efficient}, allows us to precisely slice the baseband data as a function of frequency to follow the dispersive sweep of the FRB on a channel-by-channel basis. This is critical at the low frequencies where CHIME operates (400-\SI{800}{\mega\hertz}).

CHIME already has a well-defined data format built on the Hierarchical Data Format (\hdf) scheme, and memory mapping tools provided by \texttt{caput}~\citep{caput} which facilitates slicing and transposing the data along various axes for high-level data management, and distributing the computations across multiple compute nodes. In this~\texttt{BBData}~format, our data volume is roughly \SI{50}{\giga\byte} per burst.
In contrast, a dump of the dispersion sweep including the whole CHIME band ($\lesssim\SI{20}{\second}$) for the full array would result in roughly \SI{20}{\tera\byte} of data per FRB. On the basis of the mature software tools readily available within CHIME for processing in our native format, we determined that re-formatting to match the assumptions of commonly-used formats would be impractical. For instance, if we were to use
~\vdif~\citep{whitney2010vlbi}, the fixed size of the header would be cumbersome, and the ~\vdif~assumption that time is the
slowest-varying index in our native format, would require grouping our channels into sub-bands, adding varying amounts of zero padding in each channel, and transposing the time and frequency axis--an operation we deemed too cumbersome given our current tools.

After correlation, the visibilities produced by~\corrname~are stored in a file format based on the \hdf~framework, which can be readily accessed via our package~\coda.~\coda~includes tools to solve for and apply phase/delay/rate/ionospheric corrections, station-based clock corrections, and finally a single-pulse VLBI localization pipeline. Unlike our very specific baseband data format, our visibility products are more conventional; in future work we may standardize our visibilities into a
common data format (e.g.
CASA MeasurementSet~\citep{vandiepen2015casacore}) to access different capabilities in established VLBI analysis frameworks such as AIPS~\citep{greisen2003aips} or more recently, CASA~\citep{mcmullin2007casa,casa2022casa,vanbemmel2022casa}.

\subsection{Delay Model}
Before the data are correlated, the VLBI correlator must compensate for known delays and, if necessary, gate the data. The delay compensation is done as a function of time for a fixed set of celestial (RA, declination) coordinates often referred to as the phase center. However, we refer to it hereafter as a \textit{pointing center} to avoid confusing it for the centroid of the telescope's antenna elements, which is sometimes referred to in the electrical engineering literature as an \textit{antenna} phase center (e.g.~\citep{rothacher1995determination,mitha2022sidelobe}). 

We compute geometric delays using the package \texttt{pycalc11}\footnote{https://github.com/aelanman/pycalc11}, which provides a Pythonic interface to \textsc{CALC11}, the 11th edition of the Fortran library \textsc{CALC}. \texttt{pycalc11}'s copy of the \textsc{CALC} software was derived from the available source of \texttt{difxcalc11}~\citep{gordon2016difxcalc}, the implementation of \textsc{CALC} developed for the \texttt{DiFX} correlator. More details on \texttt{pycalc11} may be found in 
\citet{lanman_pycalc11_2024}.

\textsc{CALC11} implements the Consensus VLBI delay model \citep{eubanks1991consensus}, which achieves theoretical picosecond-scale precision by taking into account effects such as relative station motion, Shapiro delays due to massive solar system bodies, and gravitational time delay due to the Earth. Positions and velocities of solar system bodies are obtained using the JPL DE421 ephemeris~\citep{folkner2009planetary}. Earth orientation parameters (in \texttt{pycalc11}) and time conversions
are source from the IERS using \texttt{astropy} \citep{astropy:2022}. \textsc{CALC11} also corrects station coordinates for various tidal effects and adds corrections for the wet and dry troposphere components.

One important consideration relates to how \textsc{CALC11} handles antenna motion. In calculating the delay $\tau^{AB}$ between antennas $A$ and $B$, \textsc{CALC11} incorporates the motion of antenna $B$ relative to $A$. The resulting delay is then defined in the reference frame comoving with station $A$. To ensure all delays are calculated in a common reference frame, we typically use the \emph{geocentric delay} of each station, the delay between the station position and the geocenter, evaluated at the time $t^A$ at which the data are recorded; we refer to this as $\tau^{AC}(t^A)$.

\section{Chunking and applying delays}\label{sec:chunking}
%
The large volumes of baseband data in VLBI require that the data be broken up into chunks to fit the data in memory for processing. After all of the data are chunked at the top level, the delay compensation and correlation can be processed in parallel over frequencies, pointing centers, time chunks, and all four combinations of polarization pairs. Here we discuss the chunking scheme that is applied within~\corrname, which is designed such
that the low-level core of the software can be written agnostic of correlation mode, e.g. whether we are using multiple phase centers, dedispersion, or pulsar gating, which are all special cases of our generic bookkeeping scheme. Note that a table of indices and definitions is available in Appendix~\ref{tab:definitions} for convenience.


The beamformed baseband data at station $S$ constitute a three-dimensional array $B^S_{kbm}$. From slowest to fastest varying indices in memory/on disk, $k$ indexes frequency channels, $b$ indexes the sky pointing\footnote{Since CHIME is a compact interferometer whose feeds have small collecting area, we typically combine groups of antennas by phasing up the entire group towards one (or multiple) sky directions, or pointings, prior to VLBI correlation.} and polarization, and $m$ is the time axis
in units of frames (note that Table~\ref{tab:definitions} contains a list of symbols used in this paper). With this data ordering, we read in a few frequencies at a time, and perform two nested loops: an outer correlation loop over all correlator pointings within that beamformer pointing,
and an inner loop for all time segments and polarization pairs within each correlator pointing. We define a \textit{scan} of data to be a time-contiguous subset of baseband data $B^S_{kbm}$ which is small enough to fit in memory and holds data for a single frequency channel (a fixed value of $k$) and beamformer pointing (a fixed value of $b$), both beam polarizations, and many time samples ($m$ values). Each scan, indexed by $n$, is further divided into subintegrations along the time axis
such that each
subintegration consists of sufficiently few time samples ($m$ values) to accurately apply delay compensation (see equation~\ref{eq:rot_timescale}).

To fully specify the correlation job we define $t^{C}_{kpn}$, the absolute start time of each integration in the geocenter frame (indicated by a $C$), in frequency channel $k$, for pointing $p$, and as a function of scan number $n$. The duration of the scan in the geocenter frame is some total width $w_{kpn}$.

Along the frequency direction, $t^C_{kpn}$ is allowed to vary arbitrarily. This adds significant flexibility to the correlator: for example, it lets us specify frequency-dependent gate offsets in the correlation to follow dispersive sweeps by varying the start time as a function of $k$. Along the time axis, $t^C_{kpn}$ must be monotonically increasing (i.e. scans corresponding to larger $n$ are later in time than previous ones). We can support pulsar gating mode with on- and off- gates defined
by intricate pulsar timing models, by adding offsets along the scan ($n$) axis. This can be done independently for each pointing $p$, allowing for very flexible observation modes, e.g. searching for faint in-beam calibrators or compact persistent radio emission near a fast transient. 

Since we desire sub-nanosecond delay precision by synthesizing the whole band, the absolute precision of $t^C_{kpn}$ needs to be specified to within the inverse bandwidth of the telescope to line up the data streams, even though each individual channel has a time resolution of $\Delta t=\SI{2.56}{\micro\second}$. The start of the observation at station $S$ is given by

\begin{equation} 
    t^S_{kpn} = t^C_{kpn} + \tau^{SC}(t^C_{kpn})\label{eq:station_delay}.
\end{equation}

Once $t^S_{kpn}$ is defined, then for each station $S$ we read in a scan of data between $t^S_{kpn}$ and $t^S_{kpn} + w_{kpn}$. Then, we apply delays to the scan towards each pointing.

\subsection{Delay Compensation}\label{sec:delay_comp}
A na\"{i}ve approach to delay compensation is to simply translate the entire chunk of data in time. However, this does not work because the total geometric delay varies over time; we would eventually suffer from decorrelation since the two timestreams would misalign. To avoid this, we break each scan into short \emph{sub-integrations} whose duration $\Delta$ is defined such that the delay does not change by more than $\Delta t / 10$ over the course of the scan. The maximum time between delay updates is
\begin{equation}
    \Delta \leq \dfrac{1}{10}\dfrac{c\Delta t}{2 v_{eq}} = \SI{412}{\milli\second} \times \left(\dfrac{\Delta t}{\SI{2.56}{\micro\second}}\right) \left(\dfrac{\SI{430}{\meter\per\second}}{v_{Eq}}\right)
    \label{eq:rot_timescale}
\end{equation}
where the factor of 2 arises from using the maximum \textit{relative} velocity between stations on Earth (double the Earth's equatorial velocity), and the requirement that the misalignment not exceed a \textit{tenth} of a time sample keeps decorrelation at a manageable level (see equation 9 in~\citet{keimpema2015sfxc}). 

By default, delay compensation is done with the geocenter as the common reference frame. As mentioned earlier, we bring a subintegration of data recorded at some station $A$ to the geocenter using the geocentric delay $\tau^{AC}(t^A)$; i.e. the geocentric delay is evaluated at the time at which the wavefront arrives at station $A$. Within the array of baseband data, the appropriate time varies as a function of frequency channel $k$ and subintegration number $q$, so it is a two-dimensional array
$t^A_{kq}$. However, in the
remainder of this section where we work with a single subintegration, we will drop these subscripts, using $\tau$ and $t^A$ for short.

For each subintegration, the delays are evaluated for a few frames before and after $t^A$ as to cover the subintegration, minimizing rounding errors. These delays are applied in three steps, which are summarized in Table~\ref{tab:delay_components}. First, we translate each 
subintegration by some integer number of frames (first line of Table~\ref{tab:delay_components}). 
The second step is the correction of the fractional delay that remains after rounding the delay to the nearest sample (second line of Table~\ref{tab:delay_components}). Finally, we correct for the time-varying part of the delay within the subintegration (third line of
Table~\ref{tab:delay_components}); this is equivalent to applying a Doppler shift of the data to bring it into a consistent reference frame. 

\begin{table*}[t]
    \centering
    \begin{tabular}{l l l}
\hline 
\hline 
  & Definition & Description \\
        $\tau^0_{kq}$ & = $\Delta t \times \mathrm{round}(\tau(t^A_{kq}) / \Delta t ) $ & The delay at the start of the subintegration, rounded to the nearest $\Delta t$ (abbreviated  $\tau^0$) \\
        $\tau^1_{kq}$ & =$\tau(t^A_{kq}) - \tau^0_{kq}$ & The rounding error in the previous, abbreviated ($\tau^1$) \\
        $\tau'_{kq}(m)$ & =$\tau(t^A_{kq} + m \Delta t) - \tau^0_{kq} - \tau^1_{kq}$ & The time-varying component of the delay within the subintegration, abbreviated $\tau'(m)$. \\
\hline
\end{tabular}
\caption{The total delay applied to each subintegration can be decomposed into three parts: $\tau(t^A) =\tau^0 + \tau^1 + \tau'(m)$, which are then applied to the data from one station on each baseline -- here, station A -- to compensate for the delay.}
\label{tab:delay_components}
\end{table*}

\begin{figure}
    \centering
    \includegraphics{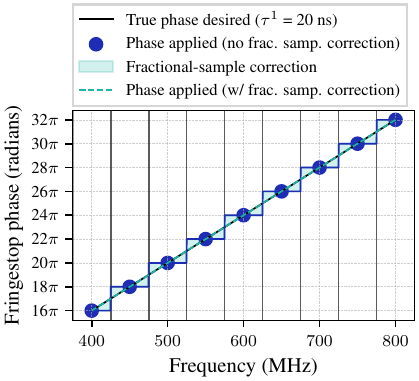}
    \caption{Illustration depicting the narrowband approximation. We plot the phase corresponding to a constant delay of \SI{20}{\nano\second} across the band. The true phase needed to compensate, or ``fringestop'', the data are shown in black. For sufficiently small delays, the narrowband approximation can be used to apply phases to the data at the central frequency of each channel (blue line and points). However, for large delays or channel bandwidths this leads to incorrect phase shift and therefore decoherence at the edge of each channel (here, we use nine channels to
    exaggerate the effect). Applying the the fractional sample correction (green) at high frequency resolution
    (equation~\ref{eq:fs_corr}) corrects this.}
    \label{fig:fs_phase}
\end{figure}

The first step of shifting the data by an integer number of frames takes care of the delay up to a rounding error. If the remaining ``fractional'' delay is small, the second step of correcting for the rounding error is straightforward. Modern connected-element radio interferometers with FX correlator backends often operate in the small-fractional-delay limit, where the delay compensation is computationally cheap. For example, CHIME's native channelization is done at a frequency
resolution of $\Delta \nu = \SI{390.625}{\kilo\hertz}$. The
largest delay expected across the physical size of CHIME is $\approx \SI{300}{\nano\second}$. If a time delay $\tau^1$ is present between two antennas within CHIME, then the residual phase gradient across the channel bandwidth is $2\pi \Delta \nu \tau^1$: this is shown in Figure~\ref{fig:fs_phase}. Requiring the residual phase to be $\leq 1$ imposes a maximum time delay of $\approx \SI{408}{\nano\second}$ across the instrument~\citep{mena2018correlator}. In this small-fractional-delay limit, we
can compensate for fractional delay efficiently by approximating the data $B_{km}$ as a single sinusoidal waveform at the center of the channel $\nu_k$. Under this ``narrow-band approximation,'' time delays (phase gradients) are equivalent to multiplicative phase shifts. Defining $\phi'_{kq} = 2\pi\nu_k\tau^1_{kq}$ and referring to the baseband data as $B_{km}$, the time shift is performed by applying a phase to the data via: 

\begin{equation}
    B_{kq} = \exp(i\phi'_{kq}) B_{kq}.
    \label{eq:no_fs_corr}
\end{equation}

The ``narrowband'' approach saves computational cost compared to performing a coherent time translation. We therefore use it at each station at the beamforming level, since the internal baselines within a station are small compared to the baselines between stations. 

\begin{figure*}
    \centering
    \includegraphics[trim = {0 4.5cm 0 0},clip]{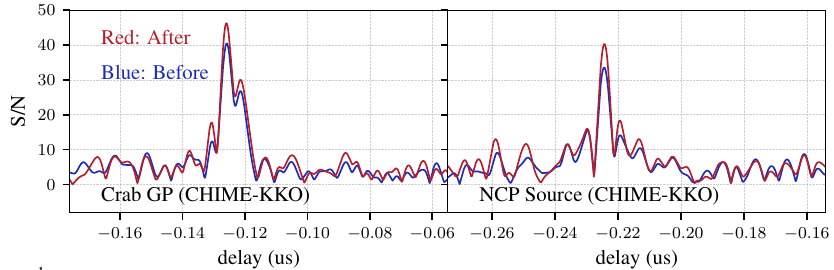}
    \caption{We show the delay autocorrelation function $G_d$ (defined formally later in equation~\ref{eq:fft_vis}) calculated from cross-correlation visibilities computed before (blue) and after (red) applying the fractional sample correction in VLBI observations on the CHIME-KKO baseline of a Crab giant pulse (left) and the NCP Source (right). This validates our implementation of the fractional sample
    correction part of our delay compensation.}
    \label{fig:fs_data}
\end{figure*}
At the VLBI level the fractional sample delay is not necessarily small (it is uniformly distributed over $[-\Delta t/2,\Delta t/2]$). To recover the misaligned signal, we apply a coherent time translation to the data. This is often referred to as the ``fractional sample correction.'' For each subintegration we apply a fractional-sample correction by Fourier transforming over the time axis, applying a phase gradient, and transforming back, before performing an overall phase shift of $\exp(2\pi i \nu_k \tau^1)$ to reflect
the fact that the carrier frequency is $\nu_k$. We implement the equation

\begin{equation}
    B_{km} \leftarrow \exp(i \phi'_{kq}) \mathcal{F}_{m\leftarrow \nu}\left( \mathcal{F}_{\nu\leftarrow m}(B_{km}) \exp(2\pi i \nu \tau^1_{kq}) \right)\label{eq:fs_corr}
\end{equation}

where the Fourier transform from the time frame axis ($m$) to the intra-channel frequency ($\nu$) axis is $\mathcal{F}_{\nu\leftarrow m}$ and its inverse as $\mathcal{F}_{m\leftarrow \nu}$. The intra-channel frequency $\nu$ is defined in terms of the total frequency of the signal $f$ via $f = \nu_k + \nu$, where $\nu_k$ is the frequency of the nearest channel center; this reflects the fact that data coming out of the telescope are already divided into many channels whose centers are $\nu_k$. 

We validate our implementation of the fractional sample correction by comparing visibilities calculated with it (equation~\ref{eq:fs_corr}) and without it (equation~\ref{eq:no_fs_corr}). In Figure~\ref{fig:fs_data}, we show visibility phases obtained in two observations: one of a single giant pulse (GP) from the Crab and the well-studied continuum source NVSS J011732+892848. Being in the VLBA Radio Fundamental Catalog, this source has compact structure and has a brightness of $\approx 5$ Janskys~\citep{condon1998nrao,yatawatta2013initial} in our band. It is located in the continuous viewing zone of CHIME
near the North Celestial Pole, making it a convenient target for long-term monitoring. We refer to this source hereafter simply as the ``NCP Source.'' In both cases the visibility phases remain largely unchanged, and the signal-to-noise ratio quantified by the FFT of the visibilities over the frequency axis improves by $10-20\%$.

Finally, after compensating for $\tau^0$ and $\tau^1$ (with either method), we need to apply a correction for the time-varying part of the delay. Note that the dominant component of $\tau'(m)$ is linearly-varying as a function of $m$ and starts at zero at $m = 0$ since the constant component has been removed, and the subintegration is short enough that higher-order terms are small, that is $\tau'(m) \approx (m\Delta t) (d\tau/dt)$. It is useful to be able to conceptualize what
it means to apply this delay in both the time and frequency domains. In the time domain, applying a changing delay to a subintegration of data can be implemented as a phase shift which varies as a function of $m$: 
 
\begin{equation}
    B_{km} \leftarrow B_{km} \exp(2\pi i \nu_k \tau'(m)) \label{eq:doppler_corr}
\end{equation}

Applying a multiplicative phase correction in the time domain which is linearly changing in time is equivalent to performing a convolution in frequency space. The convolution kernel is a delta function offset from zero intra-channel frequency; performing the convolution shifts each spectral sample to a slightly different frequency corresponding to the rest frame of the reference point (usually the geocenter). The corresponding shift in frequency is $\nu_k d\tau/dt$, which is the Doppler shift.

\section{Gating and Dedispersion}\label{sec:gating}
After performing delay compensation towards a fiducial position, the individual subintegrations are re-assembled into a scan.~\corrname~can then apply gating to remove noise-dominated samples while preserving signal-dominated samples. This is a particularly important feature for dispersed fast transients such as pulsars and FRBs. The intrinsic pulse widths of these sources can be on the order of milliseconds, but the pulse arrival time is delayed by much larger amounts at
low frequencies due to interstellar dispersion. In interstellar dispersion, the phase delay
is $\varphi(f) = \kdm \mathrm{DM} / f$ where $f$ is the physical frequency of the electromagnetic wave passing through the ISM, and where the proportionality constant $\kdm = 10^4/\SI{2.41}{\second\mega\hertz\squared\per\parsec\centi\meter\cubed}$ by convention~\citep{lorimer2004handbook, kulkarni2020dispersion}. This corresponds to a group delay which scales as $f^{-2}$
and which can be extremely large relative to the intrinsic signal duration. Dispersion measures of $\sim$\SI{3000}{\dmunit} have been observed in FRBs, corresponding to a dispersive delay of tens of seconds in the CHIME band:

\begin{equation}
    \Delta t_{0.8} - \Delta t_{0.4} = \chimedatb \left( \dfrac{\textrm{DM}}{\SI{1}{\dmunit}} \right).
\end{equation}

For dispersed transients, it is necessary to apply pulse gating as a function of frequency at high (sub-megahertz) spectral resolution, or risk losing sensitivity to dispersed signals. In~\corrname~gating is done at the native frequency resolution of the data ($\Delta f =\SI{390}{\kilo\hertz}$) by adding a delay of 
\begin{equation}
    \Delta_k = \dfrac{\kdm \textrm{DM}}{\nu_k^2} \label{eq:group_delay}
\end{equation}
to the correlator start times $t^{C}_{kpn}$. Another way of implementing an incoherent dedispersion correction uses the fact that when transformed to a consistent reference frame, the dispersive delay is identical at all stations. The visibilities are therefore insensitive to the DM used for dedispersion, since the applied delays cancel out. This means that
incoherent dedispersion corrections can be implemented \emph{after} correlation~\citep{keimpema2015sfxc}. In doing so, it is crucial to shift the timestamps and
$uv$ coordinates of the
visibilities to account for the Earth rotation during the dispersive sweep. This Earth rotation effect limits sensitivity as a function of pointing error and dispersion measure. We find that the sensitivity loss for an arcminute-scale pointing error may be relevant for the small fraction of FRBs with DM $\gtrsim \SI{2000}{\dmunit}$.

We can estimate this effect as follows. If the Earth rotation is not taken into account during the dispersive sweep the burst is shifted in time by $\Delta_k$. For a channel with central frequency $\nu_k$, the phase residual is
\begin{align} 
\varphi_k &= \nu_k (\tau_\mathrm{true} - \tau_\mathrm{applied}) = \epsilon \nu_k \Delta_k \approx \dfrac{\epsilon \kdm \textrm{DM}}{\nu_k} \label{eq:wrong_time}
    \intertext{where $\epsilon$ refers to the residual delay rate. At sufficiently high frequencies, $\varphi_k$ is small, implying that the delay rate over the dispersive sweep is unimportant. However, at high DM and low frequencies, this is no longer the case. When synthesizing a delay from many channels within a band covering $\nu_c \pm B/2$, the quadratic and higher-order contributions to the phase over the band reduce sensitivity. We calculate these contributions by Taylor expanding
    equation~\ref{eq:wrong_time} about the central frequency of the band, giving}
    \varphi_k &\approx \epsilon \kdm \textrm{DM}\left[\dfrac{1}{\nu_c} - \dfrac{(\nu_k - \nu_c)}{\nu_c^2} + \dfrac{(\nu_k - \nu_c)^2}{\nu_c^3} + \ldots \right].\label{eq:taylor_exp}
\end{align}

The first two terms in the Taylor expansion will affect the astrometry and must be compensated, but will not affect our ability to find fringes, a process we will describe in more detail in Section~\ref{sec:corr_algorithms}. The constant phase offset over the band does not affect decoherence when averaging over frequency, and because fringes are typically found using an FFT of the visibilities over frequency (see discussion surrounding equation~\ref{eq:fft_vis}), the linear term is also removed without affecting sensitivity. However, fringes will be lost if the 
quadratic term of equation~\ref{eq:taylor_exp} exceeds $2\pi$ radians:
\begin{align}
    \epsilon \kdm \textrm{DM} \dfrac{B^2}{\nu_c^3} &= 0.31 \left(\dfrac{\epsilon}{10^{-10}}\right)\left(\dfrac{\textrm{DM}}{\SI{1000}{\dmunit}}\right)\gtrsim 2\pi   
    \label{eq:fringes_decorr}
\end{align}

The delay rate on the longest baseline (CHIME-GBO) in CHIME/FRB Outriggers assuming an arcminute-level correlator pointing error is $|\epsilon| \leq 2.5 \times 10^{-10}$ over all declinations visible by CHIME. According to equation~\ref{eq:fringes_decorr}, this corresponds to a maximum DM of $\SI{2000}{\dmunit}$. For CHIME/FRB Outriggers, which capture the full bandwidth for bursts at less than \SI{1000}{\dmunit}, initial coarse pointings need to be accurate at the several arcminute level to avoid sensitivity losses
from decorrelation due to the DM sweep coupling with a pointing error.

\subsection{Coherent dedispersion}
There are circumstances in which applying gating at the native frequency resolution of the data (i.e. applying incoherent dedispersion) still results in a loss of sensitivity. This happens when the intra-channel smearing timescale (equation~\ref{eq:tau_smear}) is longer than the intrinsic duration of the signal, which commonly happens for highly-dispersed FRBs and pulsars or coarse frequency resolution. In that regime, even the signal is intrinsically very brief in time, the gate duration is limited by the intra-channel smearing
timescale:
\begin{align}
    \tau_{smear} &= \dfrac{\kdm \mathrm{DM}\Delta \nu}{\nu_k^3} \label{eq:tau_smear} \\
    = \SI{0.75}{\milli\second}  &\left(\dfrac{\mathrm{DM}}{\SI{100}{\dmunit}} \right) \left(\dfrac{\SI{0.6}{\giga\hertz}}{\nu_k}\right)^3 \left(\dfrac{\Delta \nu}{\SI{390.625}{\kilo\hertz}}\right).
\end{align}

To apply tighter gating to the signal, we can apply coherent dedispersion to the delay-compensated data to resolve the signal at higher time resolution. Coherent dedispersion removes some or all components of the dispersive phase; we review these different components before discussing our particular implementation of the coherent dedispersion correction. 

In all cases, coherent dedispersion is done by upchannelizing the data in a channel centered at central frequency $\nu_k$, applying a transfer function $\exp(-2\pi i H(\nu,\nu_k))$, and then downchannelizing again. Note that the dedispersion kernel $H$ is by convention written not in terms of the total (sky) frequency $f$, but rather explicitly as a function of the central frequency $\nu_k$ and the intra-channel frequency $\nu$
defined earlier such that $f = \nu_k + \nu$. The total dispersive phase to be removed is proportional to $1/f$; we simply need to consider which parts of the total need to be removed. Expanding about $f = \nu_k$ we have
\begin{align}
H &= \dfrac{\kdm\mathrm{DM}}{f} = \dfrac{\kdm\mathrm{DM}}{\nu_k + \nu} \\
    &= \kdm\mathrm{DM} \left[\dfrac{1}{\nu_k} - \dfrac{\nu}{\nu_k^2} + \dfrac{\nu^2}{\nu_k^3} - \dfrac{\nu^3}{\nu_k^4} \ldots \right].
\intertext{Following the discussion in~\citet{lorimer2004handbook}, we note that the third and higher-order terms can be grouped as a geometric series; $H$ can be exactly represented as}
H(\nu_k,\nu) &= H_\varphi(\nu_k) + H_t(\nu_k,\nu) + H_s(\nu_k,\nu)
\intertext{where}
H_\varphi(\textrm{DM},\nu_k) &= \dfrac{\kdm\mathrm{DM}}{\nu_k}\label{eq:h1}\\
H_t(\textrm{DM},\nu_k,\nu) &= -\kdm\mathrm{DM}\dfrac{\nu}{\nu_k^2}\label{eq:h2}\\
H_s(\textrm{DM},\nu_k,\nu) &= \kdm\mathrm{DM}\dfrac{\nu^2}{\nu_k^2 (\nu_k + \nu)}.\label{eq:h3}
\end{align}
The three pieces of the kernel encode different operations. Note that for the $k^\mathrm{th}$ data channel, $H_\varphi$ is simply an overall phase shift. If we are recording only a single channel of data, or if we do not care about the relative phase between channels with different values of $k$, (e.g. when forming Stokes parameters from the data) $H_\varphi$ has no impact on the answer. $H_t$ is the familiar $f^{-2}$ group delay in each channel (equation~\ref{eq:group_delay}). The Fourier shift theorem means that applying $H_t$ translates the
data along the time axis, since it is linear in the ``intra-channel'' frequency $\nu$. $H_t$ is more frequently applied as ``incoherent dedispersion'': a translation of the data by some integer number of frames (i.e. the dispersive delay $\kdm\mathrm{DM}/\nu_k^2$ is rounded to the nearest integer multiple of $\Delta t$). We omit this time shift to avoid double counting the dispersive group delay, which is applied as frequency-dependent gating (see also the discussion surrounding
equation~\ref{eq:wrong_time}). 

Finally, $H_s$ is the part of the kernel which reverses the intrachannel smearing. Since $H_s = 0$ for $\nu = 0$ (corresponding to the center of the channel), the data are de-smeared by phase-shifting the different frequency components such that they are aligned with that of the central frequency of the channel.

In VLBI, if dedispersion is applied in the geocentric frame as in~\corrname, it is appropriate to use only $H_s$, omitting $H_t$ and $H_\varphi$. The reason is similar to that of applying incoherent dedispersion after correlation: it does not matter whether we apply $H_\varphi$, since it will cancel out when cross-correlation visibilities are formed. $H_t$ coherently shifts the arrival time of the data
to infinite frequency, which is undesirable for taking into account the time dependence of the Earth rotation through the dispersive sweep as discussed in the section surrounding equation~\ref{eq:fringes_decorr}. However, $H_s$ is useful in VLBI correlation when the intrachannel smearing is larger than the pulse width (true for high DMs and low frequencies) since it narrows the temporal duration of the signal without changing the arrival time of the pulse relative to the channel center ($\nu_k$). We
can narrow the correlation window after de-smearing, which rejects noise and boosts sensitivity.

To realize the potential sensitivity gain of de-smearing the data via $H_s$, the scan must have a duration $w_{kpn}$ greater than that of the intra-channel smearing, such that the sweep through the channel is fully captured. After delay compensation and dedispersion, we may then integrate over a small fraction of $w_{kpn}$, tuned to the de-smeared pulse width. We parameterize this as a ``duty cycle'' $r_{kpn}$ which is also set independently for every frequency channel, pointing, and scan number. For a
scan starting at $t^S_{kpn}$ and ending at $t^S_{kpn} + w_{kpn}$, we integrate the de-smeared data over the range $t^S_{kpn} + w_{kpn} / 2 \pm  r_{kpn} w_{kpn} / 2$. Therefore, $r_{kpn} = 1$ corresponds to all the data used in the final correlation (used e.g. for sources which do not benefit from gating after de-smearing); $r_{kpn} = 1/2$ means that half of the data are used in the final correlation.

The combination of $t^C_{kpn}$, $w_{kpn}$, and $r_{kpn}$, along with the sky coordinates (right ascension and declination) and an optional dispersion measure completely determine the VLBI correlation job. By specifying these numbers at the top level of a correlation job, we allow for frequency-dependent gating, frequency-dependent scan durations (useful for e.g. pulsars with long scattering tails), gating with a pulsar timing model, and integration over some fraction of the scan within the correlator for tight gating around highly-dispersed signals. Once $t^C_{kpn}$, $w_{kpn}$, and $r_{kpn}$ are
specified, the correlation job can be parallelized over any of these axes. For instance, parallelizing over frequency makes the most sense from an I/O point of view, since frequency is the slowest-varying axis (i.e. all time samples are contiguous in memory for one frequency channel). However, when sensitivity of the full band is needed to search for fringes within hundreds or thousands of pointings, the processing can be done in parallel over pointing, such that the full bandwidth is available
to search for fringes within each pointing.

\section{Correlation Algorithm and Delay Finding}\label{sec:corr_algorithms}
Once a scan of data is defined, broken into subintegrations, compensated for known delays, and gated, each channel of data at station A is correlated with other stations to calculate visibilities. The visibilities quantify the correlated flux as a function of angular scale and observing wavelength and are the basic observable used for imaging and astrometric VLBI. We form visibilities independently for all baselines $AB$, frequency channels $k$, pointings $p$, polarization pairs, scans $n$, and integer lags $l$; however we suppress all but the $k$ and $l$ subscripts in this section for clarity. The visibilities are 
\begin{equation}
    V_{kl} = \sum_{m \in \textrm{scan}} B^A_{mk} \overline{B^B_{m'k}}
    \label{eq:naive_corr}
\end{equation}
where $B^A_{mk},B^B_{m'k}$ refer to the delay-compensated and gated scans of baseband data at stations $A$ and $B$ respectively, and where the integer delay $l = m - m'$. We correlate by integrating over the scan defined by a start time, gate width, and duty cycle ($t^C_{kpn},w_{kpn},$ and $r_{kpn}$). We implement equation~\ref{eq:naive_corr} as a convolution over frames (m,m'), evaluated by Fourier transforming over the frame axis, multiplying, and inverse Fourier transforming over the
intra-channel frequency axis. This returns visibilities for each frequency channel as a function of integer delay $l = m - m'$. We expect $l = 0$ to be the integer delay with a strong signal, since we have already aligned the datasets in time. These ``off-lags'' remain valuable as a null test and for estimating our sensitivity, and encode information at delays larger than $\Delta t$. 

Prior to any scientific application of VLBI, fringes on the observing target must be found from the visibilities. The standard approach is to use the fact that point sources are unresolved and live at a single (or a small range of) values in delay space. To maximize the signal, we transform the visibilities into delay space. This is partially done already, since we already resolve the integer part of the delay $l$ (i.e. an integer multiple of $(\Delta t = \SI{2.56}{\micro\second})$). The
remaining sub-integer part is measured by Fourier transforming the visibilities at each integer delay over frequency ($\mathcal{F}_{d\leftarrow k}$). 
\begin{align} 
    G&(d,l) = \mathcal{F}_{d\leftarrow k}V_{kl}.\label{eq:fft_vis} \\
    \intertext{The signal strength of the correlation fringe and our estimate of the delay is defined by taking the maximum value of $G$ over all sub-integer delays $d$ after subtracting the RMS background noise power, estimated by the median of $G$. Since the signal is expected to be at small delays, in practice we fix $l = 0$ to decrease the search space and thus the trials factor. The signal-to-noise ratio ($S/N$) can then be calculated from the signal power $S$ and the noise power $N$ defined below. The signal strength is measured by taking the maximum over sub-frame delays $d$, i.e.}
    S &= \max_{d}(G(d,l=0)) - \mathrm{median}( G(d,l=0)).\label{eq:signal}
    \intertext{The noise power can be estimated using the median absolute deviation (MAD) of $G$ for $(d,l=0)$. We opt for the median absolute deviation instead of the standard deviation to make our noise estimates more robust in the presence of a signal, i.e. a peak in $G(d,l=0)$.}
    N &= \mathrm{MAD}(G(d,l=0).\label{eq:noise}
\end{align}
For sources with known positions the delay is likely to be close to zero. Conversely, for sources with poorly-determined positions, larger residual delays are possible. For this reason, 
during fringe finding it is crucial to maximize sensitivity as quantified by $S/N$ over a wide range of delays (i.e. field of view). This is an important consideration for widefield applications of VLBI such as imaging and surveys for faint, compact sources.

Performing correlations as described induces subtle sensitivity losses as a function of delay (i.e. far away from the pointing center) despite the innocent simplicity of equation~\ref{eq:naive_corr}. For instance, one effect arises from the FFT implementation of equation~\ref{eq:naive_corr} which assumes periodic boundary conditions along the time axis. For nonzero integer delay $l$, exactly $l$ time samples get ``wrapped around'' the edge of the scan and are spuriously included in the correlation. We are
mostly interested in fringes appearing at low delays $(|l| \lesssim 4$, or about \SI{10}{\micro\second}), whereas the scan is $\sim 10^3$ samples long even for a millisecond-duration scan. Therefore, the sensitivity loss from the small number of wrapped-around samples is negligible over small fields of view, but at large delays this may become an issue. 

Whereas the sensitivity loss from nonzero integer delays is negligible for the delays of interest, a larger sensitivity loss arises for nonzero sub-integer delays. This arises from the fact that the data are windowed and segmented during channelization by the CHIME PFB at the time of data collection instead of at the time of VLBI correlation. In the remainder of this section we characterize this loss and develop a way to mitigate it.  

We begin with an intuitive explanation for why signal is lost. During fringe finding, we choose an integer delay ($l = 0$ in equation~\ref{eq:signal}-\ref{eq:noise} above) and use the FFT of the visibilities to find a delay where a strong peak appears. The signal-to-noise ratio will be high if all of the signal is concentrated at a single integer delay $l$; however, if there is a large uncompensated delay, the signal might lie between two neighboring integer delays. Since we
selected the zero-lag visibilities,
there is a corresponding loss of sensitivity if the signal is in actuality closer to $l = \pm 1$. We then expect our sensitivity to be highest near zero sub-integer delay and lowest when the signal is halfway between consecutive integer delays.

\begin{figure}[ht]
    \includegraphics[width=0.47\textwidth]{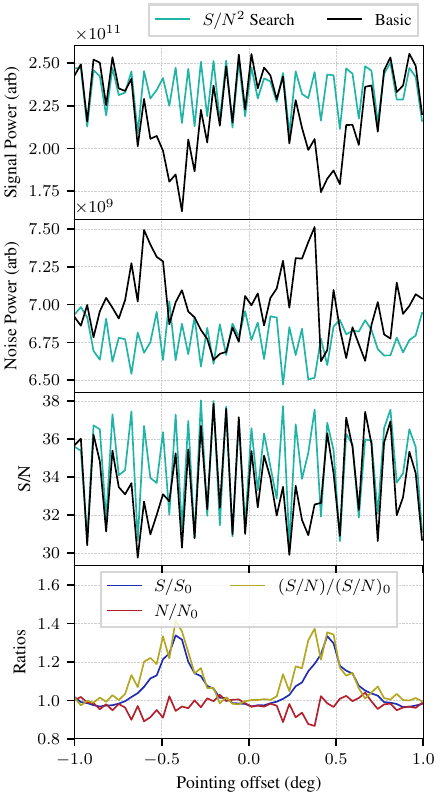}
    \caption{Top three panels: the signal power $S$, noise RMS $N$, and the resulting signal-to-noise ratio ($S/N$) recovered as a function of correlator pointing for pulsar B2310+42 calibrated to J0117+8928. With the basic correlator (black curves), the sensitivity drops significantly at around $\pm \SI{20}{\arcmin}$, which corresponds to a half-integer residual delay of $\Delta t / 2$. Using the $S/N^2$ search correlator (green;
    equation~\ref{eq:snr_correlator}), the sensitivity can be recovered near the integer lag boundaries. Bottom panel: We plot the signal and noise power ratios $(S/S_0)$ and $N/N_0$ as well as the S/N ratio ($(S/N) / (S/N)_0$) to reduce the pointing-to-pointing sample variance in the top three panels, finding a $\approx 30\%$ improvement at half-integer delays fully consistent with our simulations. This demonstrates our ability to mitigate the sensitivity losses arising from upstream PFB
    windowing effects.}
    \label{fig:pulsar_oqe}
\end{figure}

We illustrate this empirically using the 66-kilometer CHIME|KKO baseline. We correlate baseband data from CHIME and KKO pointed towards the pulsar PSR B2310+42. We apply delay solutions from the in-beam NCP source, and FFT the visibilities to measure a correlation
signal-to-noise according to equation~\ref{eq:signal}-\ref{eq:noise}. Then we repeat the process, introducing an artificial correlator pointing offset by up to $\approx 1$ degree from the true source position. This range of pointings is sufficiently wide to allow us to cover the full range of sub integer delays ($\pm \Delta t$).

The measured signal power, noise power, and signal-to-noise ratio are plotted using black traces in Figure~\ref{fig:pulsar_oqe} for the basic correlator (the other curves allow comparison of different algorithms, which we discuss later). Starting near zero pointing offset, we observe that the signal power is highest near the true pulsar position as expected. Moving farther away from zero pointing offset, the signal and noise power are both modulated; the signal power reaches a minimum at a pointing offset corresponding to a half-integer signal delay of $\pm\SI{1.28}{\micro\second}$. For
larger pointing offsets the signal strength increases again, as the fringe appears in lags $l = \pm 1$.

We see that for finding fringes on fast transients on the short CHIME-KKO baseline, the basic correlator can find fringes on sources which are localized only modestly ($\lesssim 0.5^\circ$) about their true positions. The typical pointing offset we expect in CHIME is $\approx 1'$; in this regime the sensitivity loss is negligible, justifying the use of the basic correlator (equation~\ref{eq:naive_corr}). However, on longer baselines (up to $50\times$ longer than the
CHIME-KKO baseline), poorly-localized sources may suffer sensitivity losses with correspondingly smaller position offsets.

We have developed a drop-in replacement
for equation~\ref{eq:naive_corr} which recovers full and uniform sensitivity at the price of increased correlation cost. This ``search'' correlator algorithm is derived and tested on simulations below. The search correlator achieves a $\approx 30\%$ signal-to-noise improvement, as quantified by equation~\ref{eq:signal} and~\ref{eq:noise}, on both real and simulated data and is implemented as an option in~\corrname. 
The search correlator is the last in a series of variants of equation~\ref{eq:naive_corr}, which we call the $1/N^2$ correlator, the $S/N^2$ correlator, and the $S/N^2$ search correlator. These algorithms, derived using the ``quadratic estimator'' formalism commonly used in cosmological data
analysis, take into account different types of time-domain correlations within the data to improve sensitivity. These correlations originate in the upstream windowing and channelization, which is performed by a PFB; note that FFT channelization is a special case of critically-sampled PFB channelization so it is also handled by our algorithms.

We provide a brief introduction to PFB channelization and the quadratic estimator technique in \S~\ref{sec:pfb} and~\ref{sec:oqe} respectively, and refer the reader to classic texts like~\citet{price2016spectrometers} and~\citet{tegmark1997how} for more comprehensive reviews. We perform simulations showing that these drop-in replacements improve sensitivity at the edge of the field of view (\S~\ref{sec:sims}) by an amount consistent with what we observe in real data.

\subsection{Polyphase filter banks}\label{sec:pfb}
The PFB is a generalization of a Fourier transform which can reduce spectral leakage to arbitrarily low levels at the cost of increasing temporal leakage. It splits an original signal into
$N$ frequency channels, each of which individually has a time resolution reduced by a factor of $2N$. Like Fourier transforms, PFBs may take either real or complex input voltages; the CHIME PFB takes (real-valued) voltage data (sampled at 800 MHz), and channelizes it into (complex-valued) baseband data with moderate time and frequency resolution covering the second Nyquist zone (400-800 MHz). A natural way to motivate the PFB is by considering the short-time Fourier transform (STFT)
method, one of the simplest methods for time-frequency channelization. In the STFT method, the input data are grouped into segments, or ``frames,'' of length $2N$, and for each frame of data, we perform a real to complex FFT, which returns $N + 1$ spectral channels. The Nyquist frequency can be kept but we discard it in CHIME, leaving $N$ channels. 

This can be represented mathematically as the following linear operation, which transforms unchannelized voltage data $v[j]$ into channelized baseband data $B_{mk}$. Here, $j$ indexes the time axis, $k = 0,1,\ldots 1023$ indexes our frequency channels in units of $\SI{390.625}{\kilo\hertz}$, and $m$ indexes frames: 

\begin{equation}
    B_{mk} = \sum_{j = 0}^{\infty} W[j - 2Nm] v[j] \exp(2\pi i j k / (2N)). \label{eq:pfb}
\end{equation}
There is some flexibility in the choice of window function $W$. For the short-time Fourier Transform described above the window function is
\begin{equation} 
W_{STFT}[j] = \begin{cases} 1 & 0 \leq j < 2N \\ 0, & \mathrm{otherwise.} \end{cases}
\end{equation}
One significant drawback of the short-time Fourier Transform (STFT) is that it induces spectral leakage between frequency channels due to the size of the FFT block (effectively a rectangle window in the time domain). The idea of the PFB is to reduce spectral leakage by extending the support of $W[j]$ beyond $0 \leq j < 2N$. In CHIME, we choose to extend the PFB window by a factor of 4 (often referred to as having a 4-tap PFB). The raw voltages are sampled from the sky at a rate of 800 Msps, or once every
\SI{1.25}{\nano\second}, and after omitting the Nyquist frequency we produce $N = 1024$ frequency channels, which corresponds to each channel having a time resolution of $\SI{2.56}{\micro\second}$. The Fourier transform size is $8N = 8192$, which produces $4096$ channels. Then we select every fourth frequency channel which suppresses spectral leakage into neighboring channels. The result is that we have less spectral leakage between neighboring frequency channels but significant amount of leakage
between neighboring time samples, since
the data at frame index $m$ becomes mixed with data from indices between $m-3,m+3$. In the CHIME PFB we use
\begin{equation} 
    W[j] = \begin{cases} \sin^2\left(\dfrac{\pi j}{8N-1}\right)j_0\left(\dfrac{\pi(j-4N)}{2N}\right) & 0 \leq j < 8N \\
    0 & \mathrm{else} \end{cases} \label{eq:chime_pfb}
\end{equation}
Certain window functions such as the STFT are completely lossless in the sense that the channelization can be perfectly inverted (neglecting quantization effects). If the channelization is invertible, the data can in principle be de-channelized and re-channelized to an arbitrary level to reduce the loss due to applying the windowing by $W[j]$. In practice, however, the real-time processing pipelines of radio telescopes make them difficult to invert robustly, or the PFB window functions themselves are fundamentally lossy and not perfectly invertible.

Nevertheless, some methods of PFB inversion have been developed, which we mention here briefly for completeness.~\citet{mcsweeney2020mwa} perform correlation with a synthesis filter technique, designing a $N$-tap long inverse filter for the PFB with the property that when it is convolved with channelized data and Fourier transformed, it de-channelizes the data and synthesizes an accurate reconstruction of the original timestream. However, perfect reconstruction requires a synthesis filter as
large as the original data stream, which is often computationally infeasible.~\citet{morrison2020performance}~perform correlation on an oversampled PFB (one where the channels are not decimated at the end of the FFT) by going into frequency space, and extracting the central portion of each PFB channel where the spectral response is most uniform. The extracted subbands are concatenated together, and the lag correlation function is
calculated by Fourier transform. This approach works well for oversampled PFBs, where all the information is preserved via the extra channels; unfortunately it is not applicable for critically-sampled PFBs in which the channels are decimated and do not overlap (hence losing redundancy).

Another way of illustrating the difficulty of PFB inversion is with the following example. Consider the voltage waveform corresponding to the Dirac comb ($v[j] = \sum_{m} \delta(j - 2mN)$), which gets mapped to zero by equation~\ref{eq:pfb} with $W[j]$ defined as in equation~\ref{eq:chime_pfb}. When inverting the PFB, the Dirac comb and other singular modes of the PFB need to be appropriately identified. However, the exact form of these singular modes is distorted by omitting data from the reconstruction, e.g. frequency channels saturated
(and irreversibly corrupted) by RFI. The inversion process in the presence of these practical considerations is therefore nontrivial and calls into question whether artifacts in the reconstructed timestream may be introduced by the regularization. These factors pose a considerable problem for PFB inversion, so we proceed by designing our correlation algorithm assuming that ``going backwards'' by de-channelization and re-channelization is not possible. We instead model these correlations
in the correlation algorithm.

\subsection{Quadratic Estimators}\label{sec:oqe}
One way of accounting for the windowing losses is using the quadratic estimator formalism, which is commonly used for cosmological data analysis, e.g. measuring angular power spectra of the cosmic microwave background (CMB). This approach is motivated by the close analogies between CMB data analysis and radio interferometry. In CMB analysis, the input data are a sky map comprised of pixels, which are assumed to have Gaussian noise fluctuations after foreground masking. The data $x_i$ are compressed into a summary statistic, such
as the angular power spectrum $C_\ell$ as a function of
angular frequency, which contains the information about cosmological parameters. In radio interferometry we use input data
(PFB-channelized baseband data, which have reasonably Gaussian noise properties once RFI is sufficiently removed), compress it into a summary statistic (visibilities for each channel frequency), and then extract observables such as delays and phases.

The optimal quadratic estimator formalism uses the fact that the data only have Gaussian noise fluctuations. One way of saying this is that all third and higher-order moments of the data vanish, while the second moment of the data $\langle x_i x_j \rangle$ completely describes its statistical properties through the expectation value of the two-point function.
\begin{align}
    \langle x_i x_j \rangle &= \sum_{\ell} C_{\ell} P^\ell_{ij}\label{eq:P_ell_def}
    \intertext{where $C_\ell$'s are the unknown values of the summary statistic, and $P^\ell_{ij}$ are the derivatives of the two-point function with respect to the summary statistic, which can be analytically calculated. In this case, the optimal estimator for the $C_\ell$ is given by equation 23 in~\citet{tegmark1997how}:}
    \hat C_\ell = \dfrac{1}{2}F^{-1}_{\ell \ell'} [x_i (C^{-1})_{ik} &P^{\ell'}_{km} (C^{-1})_{mj} x_j]\label{eq:recipe} 
\end{align}

It can be useful to clarify how equation~\ref{eq:recipe} translates into procedural steps. It roughly says that each data vector $x_i$ should be inverse-covariance weighted with $C^{-1}$. After this, the matrix $P_{km}^{\ell'}$ implements the reduction of the data. Finally the Fisher matrix $F^{-1}$ deconvolves artifacts at other multipoles $\ell$ arising from each multipole $\ell'$; this gives the cleaned summary statistic $\hat{C}_\ell$. 

We apply this idea to correlating voltages where the summary statistic is the time-delay correlation function ($G^{AB}_d$) between two stations $A$ and $B$ as a function of sub-integer delay $d$; for simplicity we assume that the integer delay has already been compensated so that we can always work with the $l = 0$ case. The optimal quadratic estimator will be of the form 

$$\hat G^{AB}_d = \dfrac{1}{2}F_{dd'}^{-1}[\hat{B}^A_{mk} P^{d'}_{mk~m'k'} \overline{\hat{B}^B_{m'k'}}]$$
where the hat over each dataset $B_{mk}$ indicates that the data have been weighted according to their inverse covariance. 

We now begin by calculating $P^d_{mk~m'k'}$. It is defined as the derivative of the two-point function of the data $\langle B_{mk} \overline{B_{m'k'}} \rangle$ with respect to the summary statistic, which in our case is the correlation function $G^{AB}_d$ as a function of delay. Since 
\begin{align}
    P^d_{mk~m'k'} &\equiv \dfrac{d}{dG^{AB}_d}\left( \langle B_{mk}^A \overline{B}_{m'k'}^B\rangle\right),\\
    \intertext{we have}
    \langle B^A_{mk} \overline{B^B_{m' k'}} \rangle &= \sum_d G^{AB}_d P^d_{mk~m'k'}
\end{align}
such that when there is no correlated flux ($G^{AB}_d = 0$) the expectation value of the two-point function vanishes. Expanding the two point function of the data in terms of the voltage correlation function using the definition of the PFB in equation~\ref{eq:chime_pfb} gives
\begin{widetext}
\begin{align}
    \langle B^A_{mk} \overline{B^B_{m' k'}} \rangle &= \sum_{j,j' = 0,0}^{\infty,\infty} W[j - 2Nm] W[j' - 2Nm'] \exp(2\pi i (j k - j' k') / (2N)) \langle v[j] v[j'] \rangle.\label{eq:twopf}
    \intertext{Since $W[j]$ is nonzero over $8N$ time samples, the sums here correspond to summing over a square in $j,j'$ space with side length $8N$. We can change variables to a diagonal/anti-diagonal coordinate system $a$ and $d$ such that the sub-frame delay is precisely $d$. In these coordinates we have $j = \dfrac{a + d}{2}$, $j' = \dfrac{a - d}{2}$, and}
    G^{AB}_d &= \left\langle v^A\left[\dfrac{a + d}{2}\right] v^B\left[\dfrac{a - d}{2}\right] \right\rangle
    \intertext{where we assume that the voltage timestreams $v^A$ and $v^B$ are translation-invariant and that the ensemble average $\langle \cdot \rangle$ is well-approximated by a time average over all possible $a$ values:}
    \langle B^A_{mk} \overline{B^B_{m' k'}} \rangle &= \sum_{d= - 8N}^{d = + 8N} G^{AB}_d \exp\left(\dfrac{2\pi i d (k + k')}{4N}\right) \sum_{a = d}^{a = 8N  - d} \exp\left(\dfrac{2\pi i a (k - k')}{4N}\right) W[\dfrac{a + d}{2} - Nm]W[\dfrac{a - d}{2} - Nm']. \\
\intertext{Differentiating, we obtain a formula for the two-point function derivative (equation~\ref{eq:pd_full}). It says that in the correlation process, to obtain the best estimate for the delay correlation function, we need to take into account correlations between different frequency channels (since $P^d_{mk~m'k'}$ is nonzero for $k \neq k'$):}
    P^{d}_{mk~m'k'} &= \exp\left(\dfrac{2\pi i d (k + k')}{4N}\right) \sum_{\alpha = |d|}^{\alpha = 2NM - |d|} W[\dfrac{\alpha + d}{2} - 2Nm]W[\dfrac{\alpha - d}{2} - 2Nm'] \exp\left(\dfrac{2\pi i \alpha (k - k')}{4N}\right) \label{eq:pd_full}. \\
    \intertext{The full correlation among all $N(N-1)/2$ pairs of frequency channels is not practical, but we can safely ignore correlations between neighboring frequency channels if we assume spectral leakage is subdominant to the total noise, an assumption which is valid even in real data. We write $P^{d}_{mk~m'k'} \approx P^{d}_{mk~m'k'} \delta_{k k'}$. Under this approximation, our quadratic estimator for the lag correlation
    function is merely the Fourier transform of the quadratic estimator for the visibilities. This is important because even we set out to derive the optimal method to mitigate windowing loss in the time domain, (calculating $G^{AB}_d$); in the end we get a recipe for calculating visibilities as a function of frequency channel using Eq.~\ref{eq:pd_approx} and beyond, which is much more applicable in a practical setting. This enables us to implement the quadratic estimator method
    in~\corrname~as a completely transparent replacement for the basic correlator in equation~\ref{eq:naive_corr}:}
    P^d_{mk~m'k'}&\approx \delta_{k k'}\exp\left(\dfrac{2\pi i kd }{2N}\right) \sum_{\alpha = |d|}^{\alpha = 8N - |d|} W[\dfrac{\alpha + d}{2} - 2Nm]W[\dfrac{\alpha - d}{2} - 2Nm'] \\
    &= \delta_{k k'} \exp\left(\dfrac{2\pi i kd }{2N}\right) K[d + 2N (m'  - m)]\label{eq:pd_approx}
\end{align}
\end{widetext}
where we have defined the \textit{delay}-space window function $K$, which represents the convolution of the PFB window function with itself as a function of integer delay $l = m' - m$ and sub-integer delay $d$. We re-write the function $K$ as:
\begin{equation} 
     K^d_{m m'} = K^d_{l} = K[j = d + 2N(m'-m)] = \sum_{\alpha = -\infty}^{\alpha = \infty} W[j + \alpha] W[\alpha].\label{eq:k_delay_space}
\end{equation}

We also need to approximate the covariance matrix at each station. Since $B^A_{mk}$ and $B^B_{mk}$ have zero mean, the covariance of $B^A_{mk}$ is (up to a constant)
\begin{align}
    \textrm{Cov}(B^A_{mk}) &= \langle B^A_{mk} \overline{B^A_{m'k'}} \rangle = \sum_{d} G^{AA}_d P^d_{mk~m'k'} \\
    &\approx P^0_{mk~m'k'} = K^0_{m m'} \delta_{k k'}
    \label{eq:cov}
\end{align}
under the assumption that the autocorrelation lag-correlation function $G^{AA}_d$ is dominated by system/sky noise, which shows up at $d = 0$. The last intermediate product is writing down an expression for the Fisher matrix
\begin{align}
    &F_{d d'} = \dfrac{1}{2} \mathrm{Tr}[P^{d}_{ij} (C^{-1})_{jk} \overline{P^{d'}_{kl} (C^{-1})_{lm}}] \\ 
    = &\begin{cases} \dfrac{1}{2} \sum_{l} K^d_{l + \Delta l} K^{d'}_{l} & \textrm{for }d - d' = 2N \Delta l\\ 0 & \textrm{otherwise} \end{cases} \label{eq:fisher}
\end{align}
We see that the Fisher matrix depends on the autoconvolution of the \textit{delay} space kernel and depends on the sub-integer delay $d$ and integer delay $\Delta l$. A true signal at lag $d + l\Delta t$ creates weaker artifacts at $d + (l \pm 1)\Delta t$, $d + (l \pm 2) \Delta t$, etc. 

We now lay out the qualitative steps described earlier in terms of Eqs.~\ref{eq:pd_approx},~\ref{eq:cov},~\ref{eq:fisher}. 
For starters, observe that the basic correlator in the main text can be written as a quadratic estimator as 
\begin{align}
    V_{kl} &= \sum_{m,m'} P^l_{m,m'} B^A_{mk} \overline{B^B_{m'k}}\label{eq:channelized_oqe}
    \intertext{where our choice of $P^l_{m,m'} = \delta(m-m'-l)$.}
    \intertext{To deconvolve the noise covariance, we define}
    \hat{B}_{mk}^A &= (K^0_{m m'})^{-1} B^A_{m'k} \\
    \hat{B}_{mk}^B &= (K^0_{m m'})^{-1} B^B_{m'k}, \label{eq:step1}\\
    \intertext{One could define a correlation algorithm by simply correlating the inverse noise covariance weighted data:} 
    V_{kl} &= \sum_{m,m'}\hat{B}_{mk}^A \overline{\hat{B}_{m'k}^B} \delta(m - m' - l).\label{eq:ivw_correlator}
    \intertext{We call this the $1/N^2$ correlator. However, we are not done, since the more important source of correlations arises not from the noise but from the signal itself, which correlates between neighboring frames at \textit{different} telescopes at some characteristic delay. equation~\ref{eq:pd_approx} has a copy of $K$; therefore at one station we should also convolve $\hat{B}$ with the signal kernel}
    \tilde{B}^{dA}_{m k} &= K^d_{m m'} \hat{B}^A_{m' k}~\label{eq:step2} \\
    \intertext{which defines the slightly-better $S/N^2$ correlator:}
    V_{kl} &= \sum_{m,m'}\tilde{B}_{mk}^{Ad} \overline{\hat{B}_{m'k}^B} \delta(m - m' - l).\label{eq:snr_correlator}
\end{align}
Under our reasonable assumptions of low channel leakage and input data which are system-temperature dominated and consist of signal and noise components drawn from a Gaussian distribution, equation~\ref{eq:snr_correlator} is the best correlator for extracting delays. However, it presents a chicken-and-egg problem: to do the correlation using $\tilde{B}$ requires that we already need to know \textit{a priori} the total sub-integer delay $d$ in order to calculate the best signal kernel coefficients
$K^d_{m m'}$. One way to get around this is to use the ``search $S/N^2$ correlator,'' which we implement in~\corrname~by attempting to correlate the data with a small number of trial delays $d$ (for results involving the search $S/N^2$ correlator presented here, we search over six values of d ($6d/N =  0,1,2,3,4,5$) and choose the highest signal-to-noise over the six trials. Not knowing the true delay means that at present, the search $S/N^2$ correlator makes the correlation step
a few times more costly. 

Regardless of which correlator is used, the next step is to take a Fourier transform over frequency. This arises from the fact that beyond equation~\ref{eq:pd_approx}, everything depends only on $m$ and $m'$ and is independent of $k$. What this tells us is that under the ``no spectral leakage'' approximation of equation~\ref{eq:pd_approx} the optimal quadratic estimator for the cross-correlation function comes from carefully calculating the per-frequency visibilities. By computing the visibilities as specified 
by equation~\ref{eq:snr_correlator}, we lose no information about the final answer (the delays measured by the time-lag autocorrelation). All of these correlation options are defined in~\corrname~along with an implementation of the CHIME PFB window function which can be easily swapped out for others.
If we implemented the ``optimal'' quadratic estimator, we could deconvolve our estimate of $G^{AB}_d$ by $F_{dd'}$ to remove these artifacts but if we restrict ourselves to a fixed value of the integer delay, e.g. $l = 0$, these artifacts do not affect downstream analysis, so at present, the final deconvolution is omitted.

\subsection{Benchmarking with Simulations}~\label{sec:sims}
To benchmark the new correlators, we simulate two unchannelized voltage timestreams ($v^A[j]$ and $v^B[j]$) with independent, unit-RMS Gaussian voltage fluctuations. In the ``A'' timestream we inject a signal timestream. In the ``B'' timestream we inject the same signal realization with a static delay $\tau$ relative to the ``A`` timestream. Both timestreams are
channelized with the CHIME PFB and correlated: first with the basic correlator and then with its three variations. This process is repeated for six several different sub-integer delays ($d/(2N) = 0, 1/6, 2/6, 3/6, 4/6, 5/6$) and a variety of signal strengths from below to well above the fringe detection threshold.

\begin{figure*}
    \centering
    \includegraphics[width=\textwidth]{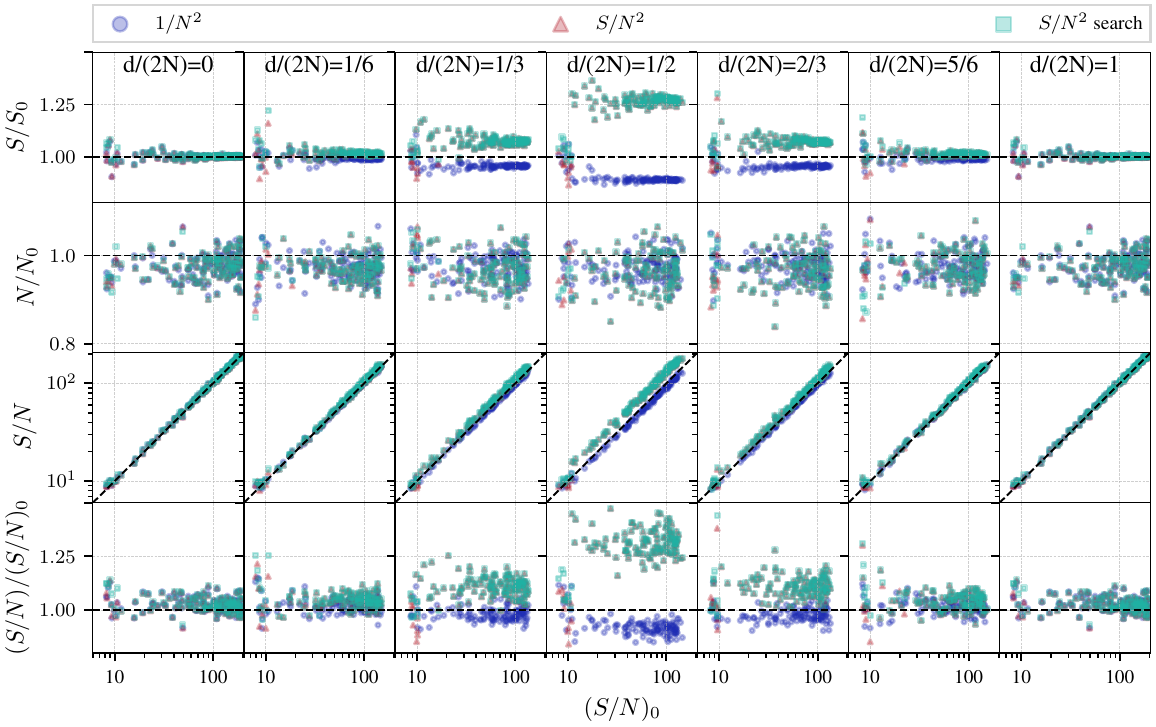}
    \caption{A realization-by-realization comparison of the signal power, noise power, and signal-to-noise ratio measured by each correlator variant (labeled SNR) as a function of the signal-to-noise measured by the basic correlator. The black lines denote equality between the basic and improved correlator variants. Each panel shows a signal injected at a different fractional sample delay $0 \leq d/(2N) \leq 1$. In most realizations of the data, the variants of the correlator improve upon the basic one, especially at half-frame lags, while the improvement is less pronounced for small delays.}
    \label{fig:sim_snr_comparison}
\end{figure*}

For each set of simulated visibilities, we calculate the detection signal power $S$ (equation~\ref{eq:signal}) and noise power $N$ (equation~\ref{eq:noise}), as a function of the sub-integer lag $\tau$ at which the signal was injected. 
Figure~\ref{fig:sim_snr_comparison} shows the performance of the other correlators. From the third and fourth row we immediately see that the signal-to-noise ratio measured by the $1/N^2$ correlator performs slightly worse than the basic correlator (black line corresponding to unity, i.e. $y=x$), but that the $S/N^2$ and $S/N^2$ search correlators perform better by $\approx 30\%$, especially at half-integer delays. 
The origin of this improvement can be analyzed in the top two rows. In the top row by plotting the ratio of the signal power ($S$) relative to signal power measured in the basic correlator $(S_0)$, we see that for the same baseband data, the $S/N^2$ and $S/N^2$ search correlators improve the signal power by $\approx 30\%$ over the basic correlator in the case of a half-integer delay ($d = N/2$).
In the second row we plot the noise statistic $N$ (equation~\ref{eq:noise}) for each correlator variant divided by the noise statistic calculated for the basic correlator ($N_0$).
After correlation, whether the basic correlator or its improved version is used to compute visibilities, we save a pre-determined number ($\approx 40$) of integer $\SI{2.56}{\micro\second}$ lags for each baseline and pointing. We choose this driven by storage limitations and the magnitude of expected clock errors.


\section{Validation of the VLBI Correlator}\label{sec:validation}
\begin{figure*}
    \centering
    \includegraphics[width = 0.98\textwidth]{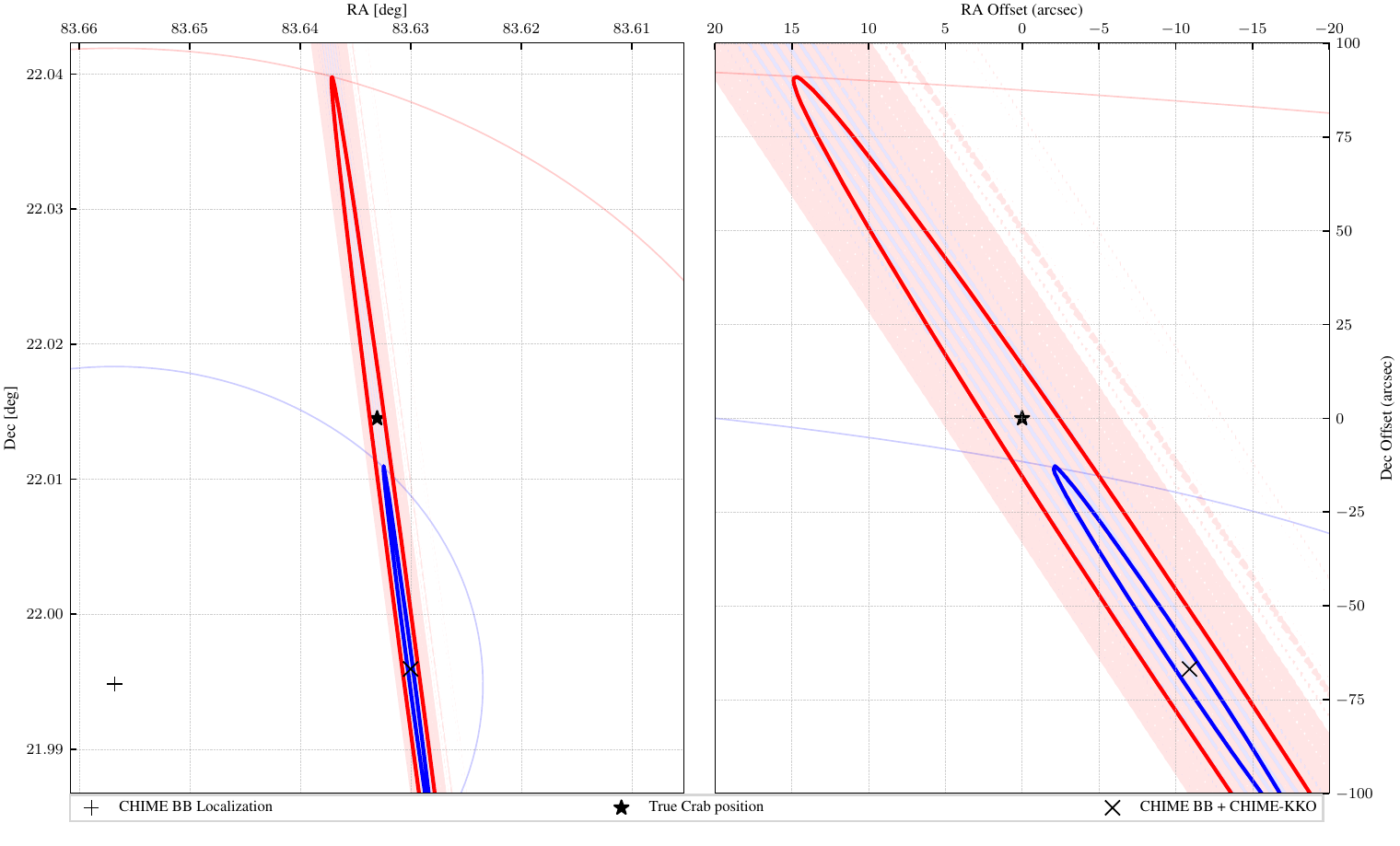}
    \caption{An example single-baseline localization of a Crab giant pulse. Left panel: a zoomed-out plot of various localization contours. A CHIME-only localization is plotted as a wide ellipse encompassing both statistical and systematic errors (light blue and red contours show the $2\sigma$ and $1\sigma$ uncertainty levels). The VLBI localization contour encompassing statistical errors only (shaded red and blue contours show the $20\sigma$ and $10\sigma$ uncertainty levels). Finally, we show the
    VLBI localization contour after taking into account the systematic uncertainty of $\sigma_\tau = \SI{2}{\nano\second}$ in equation~\ref{eq:l_tau}~(heavy red and blue lines show the $2\sigma$ and $1\sigma$ level respectively). This illustrates the relative sizes of contributions to the localization contour, which are typically dominated by systematic rather than statistical errors for single pulse observations of typical brightnesses (see~\citet{lanman2024chime} for empirical characterization of
    our systematic localization errors for the CHIME-KKO baseline).}
    \label{fig:crab_loc}
\end{figure*}

\begin{figure*}[h]
    \centering
    \includegraphics{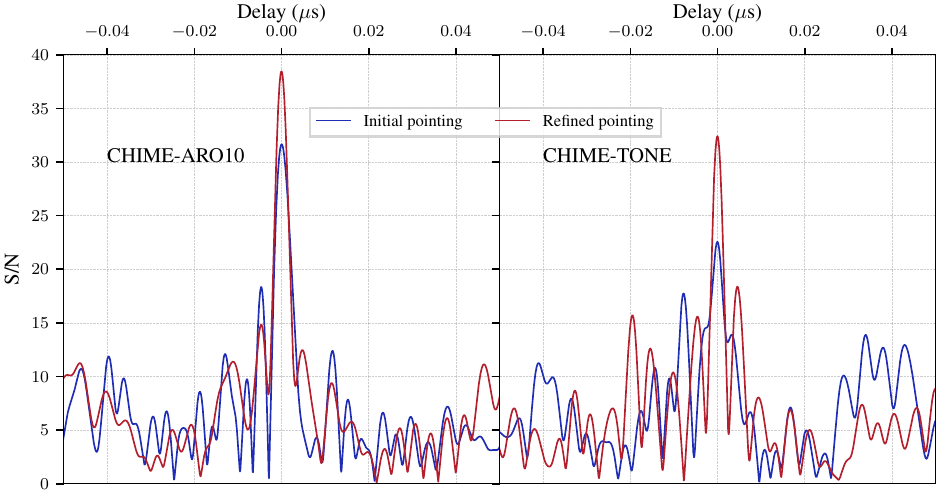}
    \caption{VLBI fringes on an FRB detected on the CHIME|ARO10 and CHIME|TONE baselines (left and right columns respectively), whose lengths are representative of those of CHIME/FRB Outriggers. This tests our delay compensation algorithms. We show the initial fringes detected (blue) while pointing at an initial guess of the FRB's position from the CHIME/FRB baseband pipeline, and the fringes after recorrelation to a refined position obtained using the method in Appendix~\ref{sec:coarse_loc},
    demonstrating the convergence of our procedure.}     
    \label{fig:FRB_20210603A}
\end{figure*}

We have already demonstrated our basic ability to find fringes in single-pulse and continuum observing modes in Figure~\ref{fig:fs_data}. As an end-to-end test of our software in the short-baseline regime we have used~\corrname~to find fringes on and localize to arcsecond accuracy over 100 single pulses from known pulsars over a wide range of signal-to-noise ratios and sky positions using CHIME and KKO; the results of that test are described in Lanman et al. (2024)~\citep{lanman2024chime} and one such localization contour is shown in Fig.~\ref{fig:crab_loc}. We have also validated~\corrname~in the regime of longer baselines and
higher dispersion measure, using our archival long-baseline data on FRB 20210603A, shown in Figure~\ref{fig:FRB_20210603A}, which we VLBI-localized~\citep{cassanelli2023fast} on continental-scale baselines using the 10-meter single dish at Algonquin Radio Observatory (ARO10)~\citep{cassanelli2022localizing} and the phased TONE array at Green Bank~\citep{sanghavi2023tone}. 

\begin{figure*}
    \centering
    \includegraphics[width=\textwidth]{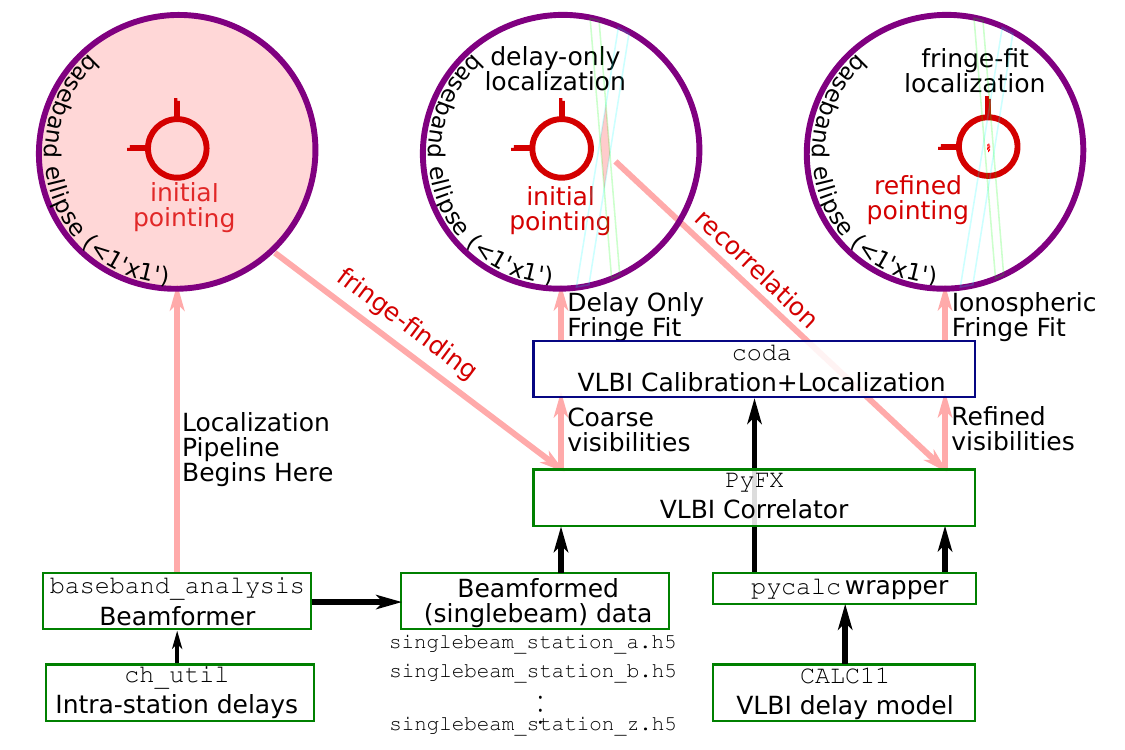}
    \caption{A high-level description of the various stages of single-pulse localization. The solid arrows denote the various stages in our pipeline. First, an initial guess of the FRB's initial position is computed, with sub-arcminute precision, from the CHIME/FRB beamformer~\protect{\citep{michilli2021analysis}}. This allows for fringes to be found, and a coarse localization within the synthesized beam refines the correlator pointing. The data are re-correlated towards the new pointing, which improves the correlation signal-to-noise.}
    \label{fig:block_diagram}
\end{figure*}

We describe the preliminary localization pipeline, depicted schematically in Figure~\ref{fig:block_diagram},
which we wrote for the localizations in both the KKO pulsar sample described and FRB 20210603A~\citep{lanman2024chime,cassanelli2023fast} with the caveat that future observations and the addition of long baselines will further drive the design of the calibration strategy. 

We begin (in the upper left circle in Figure~\ref{fig:block_diagram}) with an initial guess of the position, referred to hereafter as $\nhat_0$, which is required to be accurate at the several-arcminute level following the discussion surrounding equation~\ref{eq:fringes_decorr}. At each station, we typically form a tied-array beam in the direction of $\nhat_0$ to produce ``singlebeam'' baseband data (lower middle of Figure~\ref{fig:block_diagram}), though multi-beaming is possible.

Towards this position we find fringes using~\corrname, and use the residual group delays to localize the
source~(Appendix~\ref{sec:coarse_loc}) using $\mathcal{L}_\tau$ on each baseline (top middle of Figure~\ref{fig:block_diagram}; see also Section~\ref{sec:coarse_loc}). The correlator pointing is then updated, and the baseband data are
re-correlated towards the refined position ($\nhat_1$). The position converges within a single pointing iteration, typically with a higher S/N that that of first detection (similar to that shown in Figure~\ref{fig:FRB_20210603A}); repeated iterations do not further improve the S/N except in the case of an inaccurate initial guess.

After the coarse localization, refinement, and re-correlation, the visibilities are fringe-fitted using a model which includes positional and ionospheric phase residuals (equation~\ref{eq:phase_model}). We sample the posterior probability contours over a grid of parameters~(Appendix~\ref{sec:fine_loc}), apply as a prior the localization contour from CHIME-only baselines,
and take the maximum \emph{a posteriori} probability estimate of the pulsar's position to obtain a localization contour for the statistical uncertainties. Finally, we convolve that contour with a contour for the systematic uncertainties to obtain a final position.

\section{Discussion, Future Work, and Conclusion}
The tradeoff between efficiency and flexibility in VLBI correlation has been known for decades. We have taken flexibility to the extreme by writing an entirely Python-based VLBI correlator which supports modern features like multiple phase center correlation and incoherent \& coherent dedispersion. We believe the Pythonic implementation of our correlator is valuable and serves to open up the ``black box'' of VLBI to a larger community of astronomers, who may readily adapt it to different
telescopes or custom datasets. We have also developed and implemented within~\corrname~an algorithm
that improves the correlator's sensitivity and reduces segmentation losses from upstream channelization of voltage data: a common feature in using widefield interferometric arrays in VLBI.

\corrname~has been validated using CHIME and several outrigger telescopes, since is written with the upcoming CHIME/FRB Outriggers project in mind. In the near future,~\corrname will
soon form the basis of large numbers of VLBI observations of FRBs using CHIME/FRB Outriggers. In addition to this primary goal,~\corrname~will pave the way for ancillary scientific goals such as a low-frequency compact calibrator survey and VLBI proper motion and parallax measurements for pulsars observed by CHIME and its outrigger stations. 

More generally,~\corrname~is particularly well-suited to telescopes which channelize their data with PFBs before data recording.~\corrname~should then easily port over to several other instruments with this design, namely, CHORD, HIRAX, and BURSTT, which are currently being
constructed and plan to using single-pulse VLBI to localize FRBs.~\corrname~will also enable CHIME and its successors to join LOFAR in undertaking wide-area sky surveys~\citep{degasperin2021lofar,shimwell2019lofar,shimwell2022lofar,jackson2016lbcs,jackson2022sub,morabito2022sub} to advance VLBI at the low-frequency frontier. 

\section{Acknowledments}
\allacks
\appendix
\onecolumngrid

\section{Fringe Fitting for Point Source Localization}\label{sec:loc_methods}
After correlation and finding fringes, we typically fit the visibilities to a model for further analysis. Since fast transients are point sources, two parameters (the RA and Dec) are sufficient to describe the source structure.~\citet{nimmo2022milliarcsecond} has shown with repeated localizations of FRB 20201124A that the localization precision degrades substantially in the few-baseline limit; this will likely also impact future FRB localizations using CHIME/FRB Outriggers but depends too closely on the calibration details to conclusively determine at the time of this writing.

VLBI astrometry can be done in frequency space (using the visibilities directly) or in delay space by Fourier transforming the visibilities over the frequency axis. We use a combination of both methods. First,  a robust, delay-based localization analysis provides a coarse initial guess. Then the guess is refined by a self-consistent visibility-space analysis. We denote the complex, calibrated visibilities as $\mathcal{V}_{bk}$ where $b$ refers to the baseline, and $k$ refers to the frequency channel, and the scatter in this measurement to be isotropically distributed around $\mathcal{V}_{bk}$ in the complex plane such that the variance of each (real and imaginary) component is $\sigma^2_{bk}/2$.

\subsection{Delay-Space Localization}\label{sec:coarse_loc}
If we Fourier transform the visibilities we can directly measure the residual delay from the peak of the cross-correlation function $G_d$ in delay space. For a source whose spectrum is smooth on some characteristic frequency scale $\Delta \nu$, $G_d$ is sharply peaked at the true delay with a characteristic width $1/\Delta \nu$; this intrinsic Nyquist width is typically smaller than the systematic delay uncertainty for each baseline $\sigma_b^2$. Assuming the systematic delay errors are
Gaussian, we can therefore measure the residual delays on each baseline $\tau_{max}^b$, while taking into account the systematic delay uncertainties, by maximizing the Fourier transform of the visibilities at zero integer lag with the following expression for the log-likelihood:

\begin{equation}
    \log\mathcal{L}_\tau = \sum_b \dfrac{(\tau_{max}^b - \tau^b(\hat{\mathbf{n}}))^2}{2\sigma_b^2} \label{eq:l_tau}
\end{equation}

Fitting a position by maximizing equation~\ref{eq:l_tau} over a two-dimensional grid of sky positions is straightforward. Equation~\ref{eq:l_tau} works well when the ionospheric delay is subdominant to the geometric delay and when we are not close to the maximum lag of the FFT, such that $|\tau_{max}^b| < \Delta \tau/2$. The maximum lag of $\Delta \tau / 2$ corresponds to a
field of view of $\approx 17''$ for the longest (CHIME-GBO) baseline, and even larger for shorter baselines in the array, and ionospheric effects are important at the sub-arcsecond level, so when we are in this regime, $\mathcal{L}_\tau$ method can be used to refine the localization to the arcsecond level.

\subsection{Visibility Space Localization}\label{sec:fine_loc}
To further refine the localization to sub-arcsecond accuracy, we need to overcome the limitation of the $\mathcal{L}_\tau$ method in that it does not explicitly separate the ionosphere and geometric delays. Ideally, all contributions to the phase of the visibilities would self-consistently be taken into account and fit simultaneously to yield a best-fit (maximum likelihood) sky position.
However, a brute-force multidimensional fit is computationally expensive, so we take some steps to reduce the cost of performing the fit as follows. First, we attempt the visibility-space fit only after $\mathcal{L}_\tau$ is used to refine the localization, which reduces the search space to a few arcseconds. Second, we define our fit parameters on a per-baseline basis rather than a per-station basis, which allows the problem to be decomposed further. To illustrate this, consider a simple model
which takes into account ionospheric and geometric delays. In this case can be expressed in terms of $N+2$ parameters, with two sky coordinates and $1$ differential total electron content (TEC) 
value for each of the $N$ baselines. We refer to these parameters collectively as a vector $\vec{\lambda}$. The phase of the visibilities for a source at position $\hat{\mathbf{n}}$ and ionospheric conditions on the $b$th baseline parameterized by the differential TEC value $\rmtec_b$ is
\begin{equation} 
    P_{bk}(\vec{\lambda}) = \exp(2\pi i\nu_k \tau_{bk}(\hat{\mathbf{n}}) + i\kappa \rmtec_b / \nu_k).\label{eq:phase_model}
\end{equation}

where $\kappa = -\SI{8.45}{\giga\hertz} / \mathrm{TECU}^{-1}$, and where 1 TECU $= \SI{1e16}{\meter^{-2}}$. By parameterizing in terms of the differential TEC values per baseline, instead of the total TEC at each station, the full $N+2$ dimensional grid of parameters can be factorized into the product of $N$ three-dimensional grids: one for each baseline. For each baseline independently, we then evaluate the posterior probability of the FRB being located at each point as a function of just three parameters: the right ascension, declination, and the differential TEC on that baseline. Each of these $N$ three-dimensional grids can then be combined.

The actual value of the posterior probability computed is as follows. Following~\citet{chael2018interferometric} (see equation $16$ in that paper), we work under the assumption that the measured data are Gaussian distributed about the truth, which we model using phases $P_{bk}$ and amplitudes $A_{bk}$. The likelihood is

\begin{align}
    \log \mathcal{L}(\mathcal{V}_{bk}|A_{bk},\vec{\lambda}) &= -\dfrac{1}{2}\sum_{bk} \dfrac{|\mathcal{V}_{bk} - A_{bk} P_{bk}|^2}{\sigma_{bk}^2} \label{eq:full_logl}\\
    &= -\dfrac{1}{2}\sum_{bk} \dfrac{\mathrm{Re}[\mathcal{V}_{bk} \overline{P}_{bk} - A_{bk}]^2}{\sigma_{bk,real}^2} - \dfrac{1}{2}\sum_{bk} \dfrac{\mathrm{Im}[\mathcal{V}_{bk} \overline{P}_{bk} - A_{bk}]^2}{\sigma^2_{bk,imag}}\\
    &= -\sum_{bk} \dfrac{|\mathcal{V}_{bk}|^2}{2\sigma_{bk}^2} + \dfrac{A_{bk}^2}{2 \sigma_{bk}^2} + \dfrac{A_{bk}\mathrm{Re}[\mathcal{V}_{bk}\overline{P_{bk}}]}{\sigma_{bk}^2} \\
    \intertext{The visibility amplitudes $A_{bk}$ are nuisance parameters, and we marginalize over each $A_{bk}$ by integrating from zero to infinity numerically, assuming a flat prior on the $A_{bk}$. If we conservatively assume a uniform prior on $\vec{\lambda}$, then the likelihood becomes the posterior probability, up to a normalization constant:}
    p(\vec{\lambda}|V_{bk}) &=\prod_{bk} \int_0^\infty~dA_{bk} \exp\left(-\dfrac{|\mathcal{V}_{bk}|^2}{2\sigma_{bk}^2} - \dfrac{A_{bk}^2}{2 \sigma_{bk}^2} - \dfrac{A_{bk}\mathrm{Re}[\mathcal{V}_{bk}\overline{P_{bk}}]}{\sigma_{bk}^2} \right)\label{eq:marginalized_logl}
\end{align}
It is useful to compare equation~\ref{eq:marginalized_logl} to the standard method of producing maps from visibilities, and taking the point on the map with the maximum flux density as the most probable position. Maps are typically made by calculating
\begin{align}
    \sum_{bk} \dfrac{A_{bk}}{\sigma_{bk}^2}Re[\mathcal{V}_{bk}\overline{P}_{bk}]\label{eq:point_src_estimator}
\end{align}
(see equation 10.7 in~\citet{thompson2017interferometry1}) where the amplitudes $A_{bk}$ are fixed according to one of a number of schemes. They may be weighted neutrally ($A_{bk} = 1$), weighted as a function of baseline length (so-called ``natural'' weighting), or as a function of the signal-to-noise ratio in each channel, which upweights bright channels to increase sensitivity. When the flux from a single point source dominates the visibilities, signal-to-noise weighting corresponds to taking $A_{bk} = |\mathcal{V}_{bk}|$. In pulsar
observations, we have observed that using this weighting and taking the maximum signal-to-noise ratio as our localization estimate yields similar results as equation~\ref{eq:point_src_estimator}, though both methods (unlike the delay-based localization) are more sensitive to having accurate calibration compared to the method of equation~\ref{eq:l_tau}. Both equations~\ref{eq:full_logl} and~\ref{eq:point_src_estimator} yield statistical localization uncertainties which are much smaller than
the typical systematic uncertainties. We conclude that the choice between mapmaking and a fully Bayesian analysis is a subdominant to systematic uncertainties at present, and advocate for equation~\ref{eq:point_src_estimator} for its simplicity and robustness.

\onecolumngrid
\begin{table}[!htbp]
   \caption{A table of key variables used in this paper, including their units, properties, and typical values.}\label{tab:definitions}
   \begin{tabular}{l l}
   \hline
       $d,d'$ & Sub-integer delay, in units of the unchannelized time resolution (samples, \SI{1.25}{\nano\second}). \\
       $j, j', q'$ & Indexes time in units of samples. \\
       $k$ & Indexes frequency channels outputted by the PFB, from 0,1,...1023.\\
       $l$& Integer delay, in units of frames. \\
       $m,m'$ & Indexes time in units of the channelized time resolution (frames, \SI{2.56}{\micro\second}). \\
       $n$& Index time in units of scans (duration defined by $w$). \\
       $N$& Number of frequency channels outputted by the PFB (usually 1024). \\
       $p$ & Indexes different sky pointings within the field of view when operating in multiple phase center mode. \\
       $r$ & Indexes what fraction of the scan gets integrated into the visibilities ($0<r<1$). \\
       $t^C$ & Absolute time defined in the frame of the Earth's geocenter with a precision of a sample ($C$). \\
       $t^S$ & Absolute time defined in the frame of location $S$, e.g. topocentric time at station $A$ would be $t^A$. \\
       $\tau$& Total delay, in units of nanoseconds, before breaking into integer/sub-integer parts.\\
       $w$ & The duration of a scan, in units of the channelized time resolution.\\
       $\Delta t$ & Time resolution of channelized data (\SI{2.56}{\micro\second}). \\
       $\nu$& Intra-channel frequency, in Hz, relative to the central frequency of the nearest channel $\nu_k$; ($-\SI{195}{\kilo\hertz} < \nu < \SI{195}{\kilo\hertz}$) \\
       $\nu_k$& Central frequency in channel index $k$ ($\SI{400}{\mega\hertz} < \nu_k < \SI{800}{\mega\hertz})$. \\
       $\nu_c$& Central frequency of the entire CHIME band (\SI{600}{\mega\hertz}). \\
   \end{tabular}
\end{table}

\bibliography{all_combined_fixed_keys}
\end{document}